\documentclass[prd,floatfix,nofootinbib,eqsecnum,showpacs]{revtex4}
\usepackage{epsfig}
\usepackage{multirow}
\usepackage{amssymb,amsmath}
\newcommand{\eq}[1]{Eq.~(\ref{#1})}

\newcommand{\be}{\begin{equation}}
\newcommand{\ee}{\end{equation}}
\newcommand{\bea}{\begin{eqnarray}}
\newcommand{\eea}{\end{eqnarray}}
\newcommand{\ben}{\begin{eqnarray*}}
\newcommand{\een}{\end{eqnarray*}}

\newcommand{\DS}{Dyson-Schwinger }

\newcommand{\w}{\omega}
\newcommand{\e}{\varepsilon}
\newcommand{\al}{\alpha}

\newcommand{\ga}{\gamma}
\newcommand{\G}{\Gamma}
\newcommand{\de}{\delta}
\newcommand{\De}{\Delta}

\newcommand{\Si}{\Sigma}
\newcommand{\ro}{\rho}

\newcommand{\ka}{\kappa}
\newcommand{\ta}{\tau}

\newcommand{\pd}{\partial}

\newcommand{\cs}{{\cal S}}

\newcommand{\co}{{\cal O}}
\renewcommand{\div}{\vec{\nabla}}

\newcommand{\ov}[1]{\overline{#1}}

\newcommand{\ev}[1]{<\!\!{#1}\!\!>}

\begin{document}
\title{Quark gap equation in an external magnetic field}
\author{P.~Watson}
\author{H.~Reinhardt}
\affiliation{Institut f\"ur Theoretische Physik, Universit\"at T\"ubingen, 
Auf der Morgenstelle 14, 72076 T\"ubingen, Deutschland}
\begin{abstract}
The nonperturbative quark gap equation under the rainbow truncation and 
with two versions of a phenomenological one-gluon exchange interaction is 
studied in the presence of a uniform external magnetic field, with 
emphasis on the small field limit.  The chiral quark condensate, magnetic 
moment and susceptibility are calculated and compared to recent lattice 
data.
\end{abstract}
\pacs{12.38.Aw,11.30.Rd}
\maketitle
\section{Introduction}

Dynamical chiral symmetry breaking is one of the most important facets of 
quantum chromodynamics (QCD) in the low energy regime.  It dictates 
(amongst other things) the pattern of light meson spectroscopy and its 
continued study is of crucial relevance to our understanding of hadrons 
and their interactions.  At high energies, where the QCD dynamics are 
dominated by asymptotic freedom and the interaction is small, a massless 
(chiral) quark remains massless (and when calculated in perturbation 
theory, this applies at all orders).  However, at low energies where the 
coupling is large and perturbation theory no longer applies, chiral 
symmetry is dynamically broken and quarks attain a sizeable mass.

Alongside dynamical chiral symmetry breaking in QCD, it is also known that 
the presence of a strong constant magnetic field leads to chiral symmetry 
breaking for fermions when treated nonperturbatively 
(see Ref.~\cite{Shovkovy:2012zn} for a recent review of the topic and its 
applications).  Where dynamical chiral symmetry breaking is enhanced by 
the magnetic field, the effect is known as magnetic catalysis.  This 
effect appears even for weakly interacting fermions 
\cite{Gusynin:1994re,Gusynin:1994xp,Gusynin:1995gt}, e.g., electrons in 
quantum electrodynamics (QED).  Fermion mass generation for QED in the 
presence of magnetic fields has been studied in the context of 
nonperturbative \DS equations (see, for example, Refs.~\cite{Leung:1995mh,Lee:1997zj,Leung:2005xz,Ayala:2006sv,Rojas:2008sg}\footnote
{The reader is referred to Ref.~\cite{Shovkovy:2012zn} and 
references therein for a discussion of the many other techniques applied 
to this problem.}) using techniques based on the Ritus eigenfunction 
method \cite{Ritus:1978cj} (see also \cite{Ritus:proc}).

In this work, the nonperturbative \DS (gap) equation for strongly 
interacting quarks in an external magnetic field will be considered, with 
emphasis on the small magnetic field limit.  Unlike for electrons in QED, 
quark dynamical chiral symmetry breaking in QCD occurs (at least for 
physical values of the coupling) even in the absence of the magnetic 
field and it is of theoretical importance that the known limit of 
vanishing magnetic field be respected.  Also, the magnetic 
susceptibility, a quantity of phenomenological interest, is derived in 
the limit of vanishing magnetic field (for recent studies see, for 
example, Refs.~\cite{Frasca:2011zn,Buividovich:2009ih}).  Further, 
while there do exist physically interesting systems involving extremely 
large magnetic fields (see Ref.~\cite{Shovkovy:2012zn} and references 
therein), there are obviously many systems for which the scale of the 
strong interaction is dominant.  For example, estimates of the magnitudes 
of magnetic fields present in noncentral heavy-ion collisions 
\cite{Skokov:2009qp} indicate that they are associated with scales at 
most comparable to those of the strong interaction.  It is thus clear 
that calculations for the case of small and moderate magnetic fields 
(relative to QCD scales) should be included for a complete description 
of such systems.

As will be discussed later, the Ritus eigenfunction method, whereby an 
expansion in Landau levels is made 
\cite{Leung:1995mh,Lee:1997zj,Leung:2005xz,Ritus:1978cj}, is not 
directly applicable to the case of small magnetic fields (relative to 
the strong interaction) when applied to the nonperturbative quark gap 
equation.  Using results for the summation of Landau levels in the 
tree-level fermion two-point functions 
\cite{Gorbar:2013upa,Gorbar:2013uga}, a nonperturbative approximation 
to the quark propagator, suitable for small magnetic fields, is proposed 
and applied in this study.  The degree of dynamical chiral symmetry 
breaking is quantified by the quark condensate and this will be 
numerically evaluated in the presence of an external magnetic field.  
The magnetic moment and susceptibility 
\cite{Frasca:2011zn,Buividovich:2009ih} provide a more detailed picture 
of the system and will also be calculated.  The results will be compared 
to recent lattice data (see later for details).

The paper is organized as follows.  In the next section, the tree-level 
quark two-point functions in the presence of a magnetic field are 
discussed.  The truncated gap equation is introduced in Sec.~III.  
Section~IV then goes on to formulate the nonperturbative form of the 
quark two-point functions.  Numerical results are presented in Sec.~V.  
A summary and conclusions are given in Sec.~VI.

\section{Tree-level two-point functions}
Let us begin by laying out our conventions.  We work in Minkowski space 
(until the equations are to be numerically evaluated, whereupon we Wick 
rotate to Euclidean space) with metric 
$g^{\mu\nu}=\mbox{diag}(1,-\vec{1})$ and Dirac matrices, 
$\ga^\mu=(\ga^0,\vec{\ga})$, obeying the Clifford algebra: 
$\{\ga^\mu,\ga^\nu\}=2g^{\mu\nu}$.  It is useful to introduce the 
following (Dirac) projection matrices:
\be
\Si^\pm=\frac{1}{2}\left[\openone\pm\Si^3\right],\;\;
\Si^3=\imath\ga^1\ga^2.
\ee
Lorentz indices will be denoted with Greek symbols, $\mu,\nu,\ldots$.  
When working with the components of four-vectors, we explicitly extract 
the minus signs associated with the metric and label the components 
(excepting the Dirac matrices) with a subscript to avoid confusion with 
squares.  Further, we shall use the following notation to denote various 
collections of momentum components:
\be
p^\mu=(p_0,\vec{p}\,),\;\;
\tilde{p}^\mu=(p_0,0,p_2,p_3),\;\;
\ov{p}^\mu=(p_0,0,0,p_3),\;\;
\vec{p}_t=(p_1,p_2,0).
\ee
We shall consider quarks in the presence of a uniform magnetic field 
oriented along the $3$-axis, $\vec{B}=B\hat{e}_3$, and for clarity, 
restrict to the case where $h\equiv QB\geq0$ (with $Q$ being the 
electromagnetic charge of the quark in question).  To implement the 
magnetic field, a minimal coupling of the quark fields to the 
electromagnetic gauge potential, $A^\mu=(A^0,\vec{A})$, is made, i.e., 
$\pd_\mu\rightarrow\pd_\mu-\imath QA_\mu$ in the quark component of 
the action.  The electromagnetic gauge potential is chosen as
\be
A^0=0,\;\;\vec{A}=Bx_1\hat{e}_2.
\label{eq:pot}
\ee
With the minimal coupling, the tree-level (by which we mean free from 
QCD interactions) quark component of the action in the presence of the 
magnetic field reads
\be
\cs=\int d^4x\ov{q}_x\left[
\imath\ga^0\pd_{0x}+\imath\vec{\ga}\cdot\div_x-h\ga^2x_1-m
\right]q_x,
\ee
where ($\ov{q}$) $q_x$ is the (conjugate) quark field at position $x$ 
and $m$ is the bare quark mass.  Where possible, color and Dirac indices 
will be suppressed and we need consider only a single flavor of quark 
(later on, the approximations used in the gap equation correspond to a 
quenched system and for the numerical results, we will consider only the 
chiral case, $m=0$).  The Dirac operator may be written as
\be
D_x-m=\imath\ga^0\pd_{0x}+\imath\vec{\ga}\cdot\div_x-h\ga^2x_1-m.
\ee
The tree-level proper two-point function, $\G^{(0)}$, is defined by
\be
\G^{(0)}(x,y)=\imath\left[D_x-m\right]\de(x-y)
\ee
and the corresponding tree-level propagator, $S^{(0)}$, is the solution of
\be
\imath\left[D_x-m\right]S^{(0)}(x,y)=\de(x-y).
\ee
Beyond tree-level, the two-point functions are also related by
\be
\int d^4z\G(x,z)S(z,y)=\de(x-y).
\label{eq:leg0}
\ee

The tree-level two-point functions for a fermion in the presence of a 
uniform magnetic field were studied extensively by Ritus using 
eigenfunction methods \cite{Ritus:1978cj} (see also \cite{Ritus:proc}).  
Let us briefly review some of the pertinent results that will be of use 
in this study.  The tree-level proper two-point function can be written 
in the form
\be
\G^{(0)}(x,y)=
\sum_{n=0}^{\infty}\int\frac{d^3\tilde{p}}{(2\pi)^3}
E(x;\tilde{p},n)\G^{(0)}(\ov{p},n)\ov{E}(y;\tilde{p},n)
\label{eq:ptree0p}
\ee
where $E$ and $\ov{E}$ are the so-called Ritus matrices:
\be
E(x;\tilde{p},n)=h^{1/4}e^{-\imath\tilde{p}\cdot x}
\left[\psi_{n-1}(\e)\Si^++\psi_{n}(\e)\Si^-\right],\;\;\;\;
\ov{E}(y;\tilde{p},n)=h^{1/4}e^{\imath\tilde{p}\cdot y}
\left[\psi_{n-1}(\ta)\Si^++\psi_{n}(\ta)\Si^-\right].
\label{eq:ritus0}
\ee
In the above, the $\psi_n$ are Hermite functions 
(see Appendix~\ref{app:herm}) with discrete index $n$ and arguments
\be
\e=\sqrt{h}x_1+\frac{p_2}{\sqrt{h}},\;\;\;\;
\ta=\sqrt{h}y_1+\frac{p_2}{\sqrt{h}}.
\ee
The Ritus matrices are orthonormal and form a complete set
\be
\int d^4x\ov{E}(x;\tilde{p},n)E(x;\tilde{q},m)=
\de_{nm}(2\pi)^3\de(\tilde{p}-\tilde{q})\tilde{\openone}_n,\;\;\;\;
\sum_{n=0}^{\infty}\int\frac{d\tilde{p}}{(2\pi)^3}\,
E(x;\tilde{p},n)\ov{E}(y;\tilde{p},n)=\de(x-y),
\label{eq:proj}
\ee
where for the orthonormality, we take into account the fact that 
$\psi_{-1}=0$ with the factor
\be
\tilde{\openone}_n=
\left\{\begin{array}{cl}\Si^-,&n=0\\\openone,&n>0\end{array}.\right.
\ee
The function $\G^{(0)}(\ov{p},n)$ is the analogue of the usual momentum 
space proper two-point function and reads
\be
-\imath\G^{(0)}(\ov{p},n)=\ov{p}_{\mu}\ga^{\mu}-\sqrt{2nh}\ga^2-m.
\label{eq:tree0p}
\ee
The tree-level propagator can be written in similar fashion:
\be
S^{(0)}(x,y)=\sum_{n=0}^{\infty}\int\frac{d^3\tilde{p}}{(2\pi)^3}
E(x;\tilde{p},n)S^{(0)}(\ov{p},n)\ov{E}(y;\tilde{p},n)
\ee
with
\be
\imath S^{(0)}(\ov{p},n)=
\frac{\left[\ov{p}_{\mu}\ga^{\mu}-\sqrt{2nh}\ga^2+m\right]}
{\left[\ov{p}^2-2nh-m^2+\imath0_+\right]}.
\ee
Some explanation is in order.  As is well understood in quantum 
mechanics, the minimal coupling of a uniform magnetic field to the 
Dirac equation gives rise to discrete Landau levels for the energy 
eigenvalues (leading to the summation over the discrete index $n$ above 
and the form of the denominator factor in the propagator) with Hermite 
functions as eigenfunctions (see, for example, 
Ref.~\cite{Itzykson:1980rh}).  The characteristic combinations 
$\psi_{n-1}\Si^+$ and $\psi_{n}\Si^-$ in the Ritus matrices, 
\eq{eq:ritus0}, arise from the fact that in the Dirac equation, the 
additional term of the Dirac operator, $-h\ga^2x_1$, acts differently 
on the various spin components.  The importance of the separation of 
the $\Si^+$ and $\Si^-$ projected components of the quark propagator 
will become clear later on.

The idea behind the Ritus method is to use the (orthonormal and complete) 
matrices $E$ and $\ov{E}$, \eq{eq:ritus0}, as a substitute for the usual 
Fourier exponential factors $e^{-\imath p\cdot x}$.  After projection, 
the two momentum components of $\ov{p}$ and the index $n$ then replace 
the standard four-dimensional momentum space.  This method was applied to 
the gap equation for the two-point fermionic functions of QED 
(see e.g., Refs.~\cite{Leung:1995mh,Lee:1997zj}).

Let us now briefly review how the tree-level two-point functions reduce 
to their standard counterparts in the limit $h\rightarrow0$ 
\cite{Gorbar:2013upa,Gorbar:2013uga}.  This involves the identification 
of the Schwinger phase and the summation of the Landau levels.  Taking 
the expression for $\G^{(0)}$, \eq{eq:ptree0p}, expanding out the Ritus 
matrices and writing
\be
\ov{I}_{a,b}=\int_{-\infty}^{\infty}\frac{dp_2}{2\pi}
e^{\imath p_2(x_2-y_2)}\psi_{a}(\e)\psi_{b}(\ta),
\label{eq:int0}
\ee
one obtains the following expression:
\bea
\lefteqn{\G^{(0)}(x,y)=
\sqrt{h}\sum_{n=0}^{\infty}\int\frac{d^2\ov{p}}{(2\pi)^2}
e^{-\imath\ov{p}\cdot(x-y)}}&&\nonumber\\
&&\times\left\{\Si^+\G^{(0)}(\ov{p},n)\Si^+\ov{I}_{n-1,n-1}
+\Si^+\G^{(0)}(\ov{p},n)\Si^-\ov{I}_{n-1,n}
+\Si^-\G^{(0)}(\ov{p},n)\Si^+\ov{I}_{n,n-1}
+\Si^-\G^{(0)}(\ov{p},n)\Si^-\ov{I}_{n,n}\right\},
\eea
where with the explicit form for $\G^{(0)}(\ov{p},n)$, \eq{eq:tree0p}, 
the spin projections are given by
\bea
-\imath\Si^+\G^{(0)}(\ov{p},n)\Si^+&=&
\Si^+(\ov{p}_\mu\ga^\mu-m),\nonumber\\
-\imath\Si^+\G^{(0)}(\ov{p},n)\Si^-&=&-\sqrt{2nh}\Si^+\ga^2,\nonumber\\
-\imath\Si^-\G^{(0)}(\ov{p},n)\Si^+&=&-\sqrt{2nh}\Si^-\ga^2,\nonumber\\
-\imath\Si^-\G^{(0)}(\ov{p},n)\Si^-&=&\Si^-(\ov{p}_\mu\ga^\mu-m).
\eea
Now, using the known properties of the Hermite and Laguerre polynomials, 
$L_n^\al$, (see Appendix~\ref{app:herm}), it is possible to rewrite the 
integral $\ov{I}$, \eq{eq:int0}, in the following way 
\cite{Gorbar:2013upa,Gorbar:2013uga}:
\be
\ov{I}_{a,b}=e^{\imath\Phi(x,y)}\int\frac{d^2p_t}{(2\pi)^2}
e^{\imath\vec{p}_t\cdot(\vec{x}-\vec{y})-p_t^2/h}
\frac{2}{\sqrt{h}}(-\imath)^{a+b}\left(\frac{2}{h}\right)^{|b-a|/2}
\left\{\begin{array}{cc}
\left(\frac{a!}{b!}\right)^{1/2}(\imath p_2-p_1)^{|b-a|}
L_a^{|b-a|}\left(\frac{2}{h}p_t^2\right),&a\leq b\\
\left(\frac{b!}{a!}\right)^{1/2}(\imath p_2+p_1)^{|b-a|}
L_b^{|b-a|}\left(\frac{2}{h}p_t^2\right),&a>b\end{array}\right..
\ee
The factor $\Phi$ is the Schwinger phase and is given here by
\be
\Phi(x,y)=-\frac{h}{2}(x_2-y_2)(x_1+y_1).
\ee
As emphasized in Refs.~\cite{Gorbar:2013upa,Gorbar:2013uga}, the 
Schwinger phase encodes the information about the deviations from 
translational invariance inherent to the fermion propagator in the 
presence of the uniform magnetic field.  The remaining factors of 
$\ov{I}$ are written in the form of a standard Fourier integral in terms 
of the transverse momentum components $p_1$ and $p_2$.  Collecting the 
terms together, the tree-level proper two-point function reads
\bea
\lefteqn{-\imath\G^{(0)}(x,y)=
e^{\imath\Phi(x,y)}\int\frac{d^4p}{(2\pi)^4}
e^{-\imath p\cdot(x-y)}e^{-p_t^2/h}}&&\nonumber\\
&&\times\sum_{n=0}^{\infty}2(-1)^n\left\{
\left[-\Si^+L_{n-1}\left(\frac{2}{h}p_t^2\right)
+\Si^-L_{n}\left(\frac{2}{h}p_t^2\right)\right](\ov{p}_\mu\ga^\mu-m)
+2L_{n-1}^1\left(\frac{2}{h}p_t^2\right)\vec{p}_t\cdot\vec{\ga}\right\},
\label{eq:anolp}
\eea
where the identity
\be
\imath(\Si^+-\Si^-)\ga^2=\ga^1
\ee
is used in the last term.  Repeating the above steps for the tree-level 
propagator, one obtains
\bea
\lefteqn{\imath S^{(0)}(x,y)=
e^{\imath\Phi(x,y)}\int\frac{d^4p}{(2\pi)^4}
e^{-\imath p\cdot(x-y)}e^{-p_t^2/h}}&&\nonumber\\
&&\times\sum_{n=0}^{\infty}
\frac{2(-1)^n}{\left[\ov{p}^2-2nh-m^2+\imath0_+\right]}
\left\{\left[-\Si^+L_{n-1}\left(\frac{2}{h}p_t^2\right)
+\Si^-L_{n}\left(\frac{2}{h}p_t^2\right)\right](\ov{p}_\mu\ga^\mu+m)
+2L_{n-1}^1\left(\frac{2}{h}p_t^2\right)\vec{p}_t\cdot\vec{\ga}
\right\}.\nonumber\\
\label{eq:anols}
\eea
In the case of $\G^{(0)}$, the sum over $n$ is known 
(see Appendix~\ref{app:herm}) and is such that
\be
\sum_{n=0}^{\infty}2(-1)^nL_{n}\left(\frac{2}{h}p_t^2\right)=
\sum_{n=0}^{\infty}2(-1)^{n-1}L_{n-1}\left(\frac{2}{h}p_t^2\right)=
\sum_{n=0}^{\infty}4(-1)^{n-1}L_{n-1}^{1}\left(\frac{2}{h}p_t^2\right)=
\exp{\left\{\frac{p_t^2}{h}\right\}}.
\label{eq:nsums0}
\ee
This gives the surprisingly simple result 
\cite{Gorbar:2013upa,Gorbar:2013uga}
\be
-\imath\G^{(0)}(x,y)=e^{\imath\Phi(x,y)}\int\frac{d^4p}{(2\pi)^4}
e^{-\imath p\cdot(x-y)}\left[p_{\mu}\ga^{\mu}-m\right].
\label{eq:gtree0}
\ee
The only effect of the magnetic field on the tree-level proper two-point 
function is the introduction of the Schwinger phase.  In the limit 
$h\rightarrow0$, $\Phi\rightarrow0$ and one recovers the standard 
(translationally invariant) tree-level expression for the quark proper 
two-point function.  To obtain the propagator, one needs the following 
results \cite{Chodos:1990vv}:
\bea
e^{-\al}\sum_{n=0}^{\infty}\frac{2(-1)^n}{[\ro+2n]}L_{n}(2\al)&=&
\imath\int_{0}^{\infty}d\w
\exp{\left\{-\imath(\ro\w+\al\tan{\w})\right\}}
\left[1+\imath\tan{\w}\right],\nonumber\\
e^{-\al}\sum_{n=0}^{\infty}\frac{2(-1)^{n-1}}{[\ro+2n]}L_{n-1}(2\al)&=&
\imath\int_{0}^{\infty}d\w
\exp{\left\{-\imath(\ro\w+\al\tan{\w})\right\}}
\left[1-\imath\tan{\w}\right],\nonumber\\
e^{-\al}\sum_{n=0}^{\infty}\frac{4(-1)^{n-1}}{[\ro+2n]}L_{n-1}^1(2\al)&=&
\imath\int_{0}^{\infty}d\w
\exp{\left\{-\imath(\ro\w+\al\tan{\w})\right\}}
\left[1+\tan^2{\w}\right].
\label{eq:propint0}
\eea
With the substitutions
\be
\ro=-\frac{\ov{p}^2-m^2+\imath0_+}{h},\;\;\al=\frac{p_t^2}{h},\;\;\w=sh,
\label{eq:rodef0}
\ee
one arrives at the following parametric form for the tree-level 
propagator in the presence of the magnetic field \cite{Gorbar:2013upa}:
\bea
\imath S^{(0)}(x,y)&=&e^{\imath\Phi(x,y)}\int\frac{d^4p}{(2\pi)^4}
e^{-\imath p\cdot(x-y)}(-\imath)\int_0^{\infty}ds
\exp{\left\{\imath s(\ov{p}^2-m^2+\imath0_+)
-\imath\frac{p_t^2}{h}\tan{(sh)}\right\}}\nonumber\\
&&\times
\left\{p_{\mu}\ga^{\mu}+m+\ga^1\ga^2(\ov{p}_{\mu}\ga^{\mu}+m)\tan{(sh)}
-\vec{p}_t\cdot\vec{\ga}\tan^2{(sh)}\right\}.
\label{eq:ptree0}
\eea
In particular, for small $h$,
\be
\imath S^{(0)}(x,y)=
e^{\imath\Phi(x,y)}\int\frac{d^4p}{(2\pi)^4}
e^{-\imath p\cdot(x-y)}\left\{
\frac{[p_{\mu}\ga^{\mu}+m]}{[p^2-m^2+\imath0_+]}
+\frac{\imath h\ga^1\ga^2[\ov{p}_{\mu}\ga^{\mu}+m]}
{[p^2-m^2+\imath0_+]^2}+\co(h^2)\right\}.
\ee
The standard tree-level quark propagator thus emerges as the $h=0$ case, 
as it should.  Crucial to this is, of course, the infinite summation over 
the Landau levels -- any finite truncation of the sum would not recover 
this limit.  We notice that comparing the two-point functions in the 
presence of the magnetic field, Eqs.(\ref{eq:gtree0}) and 
(\ref{eq:ptree0}), after extracting the Schwinger phase it is not 
obvious that one is the inverse of the other and satisfy \eq{eq:leg0}.  
This means that nonperturbatively, it will prove more expedient to use 
the Ritus decomposition to connect the propagator to the proper two-point 
function and then sum (albeit under approximation).  We shall investigate 
this shortly.

\section{Truncated gap equation}
Let us now introduce the rainbow-truncated gap equation along with the 
phenomenological form for the gluon interaction and show how the Schwinger 
phase factorizes nonperturbatively.  The rainbow-truncated gap (or quark 
Dyson-Schwinger) equation is characterized by the replacement of the 
fully dressed quark-gluon vertex occurring in the nonperturbative 
self-energy integral with its tree-level counterpart, while the quark 
propagator and its inverse are dynamically dressed.  The equation reads 
(in configuration space and after resolving the color factors)
\be
\G(x,y)=\G^{(0)}(x,y)+g^2C_F\ga^{\mu}S(x,y)\ga^{\ka}W_{\ka\mu}(y,x),
\label{eq:gapp}
\ee
where $g$ is the QCD coupling and $C_F=4/3$ is the color factor 
associated with $N_c=3$ colors.  $W$ is the dressed gluon propagator for 
which we take the following Landau gauge form
\be
\imath W_{\ka\mu}(y,x)=\int\frac{d^4q}{(2\pi)^4}
e^{-\imath q\cdot(y-x)}\frac{G(q^2)}{q^2}t_{\ka\mu}(q),
\ee
where $G$ is the gluon dressing function and 
$t_{\ka\mu}(q)=g_{\ka\mu}-q_{\ka}q_{\mu}/q^2$ is the transverse 
momentum projector.  Later on, we shall use the following two 
phenomenological forms for the gluon dressing 
(with two parameters $\w$ and $d$):
\be
g^2\frac{G(q^2)}{q^2}=4\pi^2d\exp{\left\{\frac{q^2}{\w^2}\right\}}
\times\left\{
\begin{array}{cc}
\frac{q^2}{\w^2},&\mbox{type I}\\-1,&\mbox{type II}
\end{array}\right..
\label{eq:int}
\ee
The first of these forms corresponds to a simple interaction used 
previously to study dynamical chiral symmetry breaking and light meson 
phenomenology \cite{Alkofer:2002bp}.  The second is a variation of this 
(and resembles an interaction kernel constructed from lattice components 
\cite{Aguilar:2010cn}) that will provide a useful comparison.  The 
parameters $\w$ and $d$ will be chosen so as to reproduce the quark 
condensate in the absence of the magnetic field (see later).  Notice 
that the interaction is exponentially suppressed for large spacelike 
momenta and does not include the perturbative component.  At the 
technical level, this results in the simplification that one does not 
need to renormalize.  Comparing the results of 
Ref.~\cite{Alkofer:2002bp} with those of Ref.~\cite{Maris:1999nt} where 
the perturbative contributions are included, the main results are not 
significantly different, i.e., the perturbative components are not 
important when considering nonperturbative meson properties such as 
the masses and leptonic decay constants and by extension, the chiral 
condensate.

Assuming the following nonperturbative forms for the quark proper 
two-point function and propagator in the presence of a magnetic field:
\be
\G(x,y)=e^{\imath\Phi(x,y)}\int\frac{d^4p}{(2\pi)^4}
e^{-\imath p\cdot(x-y)}\ov{\G}(p),\;\;
S(x,y)=e^{\imath\Phi(x,y)}\int\frac{d^4k}{(2\pi)^4}
e^{-\imath k\cdot(x-y)}\ov{S}(k),
\label{eq:anz0}
\ee
it is clear that they reduce to the tree-level expressions, 
Eqs.(\ref{eq:gtree0}) and (\ref{eq:ptree0}).  Inserting into the gap 
equation, \eq{eq:gapp}, the Schwinger phase is simply an overall factor 
that can be removed and one obtains ($q=k-p$)
\be
\ov{\G}(p)=\imath[p_\mu\ga^\mu-m]
-\imath g^2C_F\int\frac{d^4k}{(2\pi)^4}
\frac{G(q^2)}{q^2}t_{\ka\mu}(q)\ga^\mu\ov{S}(k)\ga^\ka.
\label{eq:gap}
\ee
The absence of explicit information about the magnetic field in the gap 
equation arises because at this level of truncation, the gluon 
interaction is not coupled to the magnetic field.  The magnetic field 
dependence is in the connection between the configuration space 
functions $\G(x,y)$, $S(x,y)$ and their momentum space counterparts 
$\ov{\G}(p)$, $\ov{S}(p)$ via \eq{eq:anz0} and \eq{eq:leg0}.  Since 
\eq{eq:leg0} makes no explicit reference to the dynamical dressing of 
the functions, one can see that the Schwinger phase and the deviations 
from translational invariance inherent to the magnetic field are not 
affected by the dynamics of the theory, such that the ansatz, 
\eq{eq:anz0}, is self-consistent.  The decoupling of the deviations 
from translational invariance and the dynamics will become of 
importance below.

\section{Nonperturbative decomposition}
In order to nonperturbatively solve the gap equation in the presence of 
the magnetic field, we must decompose the quark proper two-point function 
and propagator such that \eq{eq:leg0} is fulfilled.  In the absence of 
the magnetic field, one has the standard Landau gauge expressions
\be
-\imath\G(x,y)=\int\frac{d^4p}{(2\pi)^4}e^{-\imath p\cdot(x-y)}
\left[A_L(p)p_\mu\ga^\mu-B_L(p)\right],\;\;\imath S(x,y)=
\int\frac{d^4p}{(2\pi)^4}e^{-\imath p\cdot(x-y)}
\frac{\left[A_L(p)p_\mu\ga^\mu+B_L(p)\right]}
{\left[p^2A_L(p)^2-B_L(p)^2+\imath0_+\right]}
\label{eq:lprop}
\ee
where the dressing functions $A_L$ and $B_L$ are both dependent on 
$p^2$.  At tree-level, $A_L^{(0)}(p)=1$, $B_L^{(0)}(p)=m$.  As mentioned 
previously, one might hope to take the assumed forms for the two-point 
functions in the presence of the magnetic field as given by \eq{eq:anz0} 
(where the Schwinger phase has been extracted) and use \eq{eq:leg0} to 
find one, given an explicit ansatz for the other.  However, a quick 
glance at the respective tree-level expressions, Eqs.(\ref{eq:gtree0}) 
and (\ref{eq:ptree0}), shows that this is far from trivial.  
In contrast, inserting general ans\"{a}tze for the functions with a 
Ritus decomposition into Landau levels, \eq{eq:leg0} can be resolved.

The general form for the quark proper two-point function, in terms of 
the Ritus matrices is
\be
\G(x,y)=\sum_{n,l}\int\frac{d^3\tilde{p}}{(2\pi)^3}
\frac{d^3\tilde{k}}{(2\pi)^3}
E(x;\tilde{p},n)\G(\tilde{p},\tilde{k};n,l)\ov{E}(y;\tilde{k},l).
\ee
At tree-level,
\be
-\imath\G^{(0)}(\tilde{p},\tilde{k};n,l)=
\de_{nl}(2\pi)^3\de\left(\tilde{p}-\tilde{k}\right)
\left[\ov{p}_\mu\ga^\mu-\sqrt{2nh}\ga^2-m\right].
\ee
As argued previously, the deviations from translational invariance 
encoded in the Schwinger phase factorize from the dynamical content of 
the dressed two-point function (at least under the truncation scheme 
considered here).  In terms of the above expressions, this entails that 
we may assume that the such deviations are only present in the Ritus 
matrices, which are not modified by the gluon interaction.  To ensure 
this, one may write
\be
\G(\tilde{p},\tilde{k};n,l)=\de_{nl}(2\pi)^3
\de\left(\tilde{p}-\tilde{k}\right)\G(\ov{p},n),
\label{eq:genp}
\ee
such that
\be
\G(x,y)=\sum_{n=0}^{\infty}\int\frac{d^3\tilde{p}}{(2\pi)^3}
E(x;\tilde{p},n)\G(\ov{p},n)\ov{E}(y;\tilde{p},n).
\ee
In \eq{eq:genp}, the factor $\de(\tilde{p}-\tilde{k})$ corresponds to 
momentum conservation for these components (with the electromagnetic 
potential \eq{eq:pot}, the deviations from translational invariance 
only affect the $\hat{e}_1$-components).  The restriction, $\de_{nl}$, 
on the indices is such that the integrals $\ov{I}$, \eq{eq:int0}, (from 
which the Schwinger phase emerges) are unaltered.  A suitable ansatz for 
the Dirac structure of $\G(\ov{p},n)$ is
\be
-\imath\G(\ov{p},n)=
\Si^+\ov{p}_\mu\ga^{\mu}A-\Si^+B+\Si^-\ov{p}_\mu\ga^{\mu}C-\Si^-D
-\sqrt{2nh}\ga^2E+\sqrt{2nh}[\Si^+-\Si^-]\ov{p}_\mu\ga^{\mu}\ga^2F
\label{eq:invp1}
\ee
where the dressing functions $A$-$F$ are functions of $\ov{p}^2$ and $n$ 
(where no confusion arises, we shall omit the functional dependencies of 
the dressing functions for clarity).  The above ansatz contains the 
minimal set of Dirac structures that self-consistently reproduce 
themselves when one expands the rainbow-truncated self-energy expression 
(in Landau gauge) using the techniques of Ref.~\cite{Lee:1997zj}, i.e., 
applying the Ritus eigenfunction method.  Notice that if one were to use 
the so-called lowest Landau level approximation whereby only the $n=0$ 
mode is included (as, for example, Ref.~\cite{Lee:1997zj} does), the 
characteristic Ritus matrix structures $\psi_{n-1}\Si^+$ and 
$\psi_{n}\Si^-$ reduce to the projection $\psi_0\Si^-$ and only the 
functions $C$ and $D$ in the above ansatz, \eq{eq:invp1}, contribute.  
At tree-level,
\be
A^{(0)}=C^{(0)}=E^{(0)}=1,\;\;B^{(0)}=D^{(0)}=m,\;\;F^{(0)}=0.
\ee
Making the analogous ansatz for the propagator, we have
\be
S(x,y)=\sum_{n=0}^{\infty}\int\frac{d^3\tilde{p}}{(2\pi)^3}
E(x;\tilde{p},n)S(\ov{p},n)\ov{E}(y;\tilde{p},n).
\label{eq:propa}
\ee
The Dirac structure of $S(\ov{p},n)$ in terms of the dressing functions 
is given via \eq{eq:leg0}.  Using the orthonormality and completeness 
properties, \eq{eq:proj}, of the Ritus matrices, \eq{eq:leg0} can be 
written as
\be
0=\sum_{n=0}^{\infty}\int\frac{d\tilde{p}}{(2\pi)^3}\,
E(x;\tilde{p},n)\left\{\G(\ov{p},n)S(\ov{p},n)-\openone\right\}
\ov{E}(y;\tilde{p},n).
\ee
In the above, we need not include the $\tilde{\openone}_n$ factor of the 
orthonormality relation because the functions are diagonal in the indices 
(i.e., $\sim\de_{nl}$) and we explicitly include the external projection 
matrices which reduce in the special case $n=0$.  The solution to this 
equation can readily be found and reads (the dressing functions, 
$W_A$-$W_F$, are functions of $\ov{p}^2$ and $n$)
\be
\imath S(\ov{p},n)=
\Si^+\ov{p}_\mu\ga^{\mu}W_A+\Si^+W_B+\Si^-\ov{p}_\mu\ga^{\mu}W_C
+\Si^-W_D-\sqrt{2nh}\ga^2W_E
+\sqrt{2nh}[\Si^+-\Si^-]\ov{p}_\mu\ga^{\mu}\ga^2W_F
\label{eq:prop1}
\ee
where given the combinations (we neglect the Feynman prescription for now)
\be
\De_1=\ov{p}^2AC-BD-2nhE^2-2nh\ov{p}^2F^2,\;\;
\De_2=4nhEF+AD-BC,\;\;
\De=\De_1^2-\ov{p}^2\De_2^2,
\label{eq:propden}
\ee
we have
\bea
&&W_A=\frac{\De_1C-\De_2D}{\De},\;\;
W_B=\frac{\De_1D-\ov{p}^2\De_2C}{\De},\;\;
W_C=\frac{\De_1A+\De_2B}{\De},
\nonumber\\&&
W_D=\frac{\De_1B+\ov{p}^2\De_2A}{\De},\;\;
W_E=\frac{\De_1E+\ov{p}^2\De_2F}{\De},\;\;
W_F=\frac{\De_1F+\De_2E}{\De}.
\label{eq:propfac}
\eea
We shall return to the specific form of the propagator 
(and the Feynman prescription) shortly.

With the general form for the propagator, it is now possible to 
demonstrate why we do not directly use the usual Ritus decomposition 
into Landau levels for this study.  As discussed in the introduction, 
we are interested in the small $h$ (relative to QCD scales) behavior of 
the quark gap equation.  Let us consider the quark condensate, defined 
as (trace over Dirac indices)
\be
\ev{\ov{q}q}=N_c\mbox{Tr}_{d}S(x,x).
\label{eq:cond0}
\ee
Expanding the propagator in terms of the Ritus matrices with 
\eq{eq:propa} gives
\be
\ev{\ov{q}q}=N_c\sum_{n=0}^{\infty}\int\frac{d^3\tilde{p}}{(2\pi)^3}\,
\mbox{Tr}_{d}E(x;\tilde{p},n)S(\ov{p},n)\ov{E}(x;\tilde{p},n).
\ee
Further, using the definition of the Ritus matrices, \eq{eq:ritus0}, 
we have
\be
\ev{\ov{q}q}
=N_c\sqrt{h}\sum_{n=0}^{\infty}\int\frac{d^3\tilde{p}}{(2\pi)^3}\,
\mbox{Tr}_{d}
\left[\Si^+\psi_{n-1}^{2}(\e)+\Si^-\psi_{n}^{2}(\e)\right]S(\ov{p},n).
\ee
Given that $\e=\sqrt{h}x_1+p_2/\sqrt{h}$, the integral over $p_2$ may be 
performed (it reduces to the normalization integral of the Hermite 
functions, see Appendix~\ref{app:herm}), remembering that $\psi_{-1}=0$:
\bea
\ev{\ov{q}q}&=&
N_c\frac{h}{2\pi}\sum_{n=0}^{\infty}\int\frac{d^2\ov{p}}{(2\pi)^2}d\e\,
\mbox{Tr}_{d}
\left[\Si^+\psi_{n-1}^{2}(\e)+\Si^-\psi_{n}^{2}(\e)\right]S(\ov{p},n)
\nonumber\\
&=&N_c\frac{h}{2\pi}\int\frac{d^2\ov{p}}{(2\pi)^2}\,\mbox{Tr}_{d}
\left\{\Si^-S(\ov{p},n=0)+\sum_{n=1}^{\infty}S(\ov{p},n)\right\}.
\eea
Comparing to the standard expression for the condensate in the absence of 
the magnetic field,
\be
\ev{\ov{q}q}_{h=0}=N_c\int\frac{d^4p}{(2\pi)^4}\mbox{Tr}_{d}S(p),
\ee
the differences are made clear.  The Ritus decomposition generically 
results in the replacement
\be
\int\frac{d^4p}{(2\pi)^4}\rightarrow\frac{h}{2\pi}\sum_{n=0}^{\infty}
\int\frac{d^2\ov{p}}{(2\pi)^2}
\ee
whereby the four-dimensional momentum space integration measure is 
replaced by a two-dimensional integral over the longitudinal components 
$\ov{p}$ and a sum over the Landau levels (these degrees of freedom being 
relevant for the dynamical content of the nonperturbative propagator in 
the presence of the magnetic field).  Crucially though, in order to 
maintain the dimensions, a prefactor $h$ appears.  This prefactor also 
occurs in the loop integrals associated with the self-energy when 
expanded in terms of the Ritus decomposition (see, for example, 
Refs.~\cite{Leung:1995mh,Lee:1997zj}).  Naively, all components of the 
self-energy and the condensate integrals would vanish in the limit 
$h\rightarrow0$.  Clearly, in the context of the quark gap equation 
where we know that the gluon interaction results in a nontrivial 
condensate, the Ritus decomposition must be regarded in the sense of an 
asymptotic series expansion in Landau levels labelled by $n$ (valid for 
large values of $h$).  In the limit of vanishing magnetic field, the 
Landau levels must first be summed in order to obtain a correct result.

To sum the Landau levels, it is necessary to make some assumptions and 
approximations.  The initial aim is to recover the $h\rightarrow0$ limit 
where we know that the standard Landau gauge two-point functions apply.  
In the previous section, the tree-level case was presented and will 
serve as a template for the nonperturbative case.  The first 
approximation is to set the dressing functions $F$ and $W_F$, occurring 
in Eqs.~(\ref{eq:invp1}) and (\ref{eq:prop1}), respectively, to zero:
\be
F=W_F=0.
\ee
The justification for this is that both vanish in the absence of the 
magnetic field and do not appear at tree-level.  Recalling that the 
ans\"{a}tze for $\G$ and $S$, Eqs.~(\ref{eq:invp1}) and 
(\ref{eq:prop1}), respectively, were made such that they were functions 
of $(\ov{p},n)$ so as not hinder the appearance of the Schwinger phase 
via the integrals  $\ov{I}$, \eq{eq:int0} (the interaction not 
modifying the deviations from translational invariance), we can 
immediately write down the following expressions in analogy to the 
tree-level case, Eqs.~(\ref{eq:anolp}) and (\ref{eq:anols}):
\bea
-\imath\G(x,y)&=&
e^{\imath\Phi(x,y)}\int\frac{d^4p}{(2\pi)^4}
e^{-\imath p\cdot(x-y)}e^{-p_t^2/h}\nonumber\\
&&\!\!\!\!\times\sum_{n=0}^{\infty}2(-1)^n
\left\{-\Si^+(\ov{p}_\mu\ga^{\mu}A-B)
L_{n-1}\left(\frac{2}{h}p_t^2\right)
+\Si^-(\ov{p}_\mu\ga^{\mu}C-D)
L_{n}\left(\frac{2}{h}p_t^2\right)
+2L_{n-1}^1\left(\frac{2}{h}p_t^2\right)\vec{p}_t\cdot\vec{\ga}E
\right\},\nonumber\\
\imath S(x,y)&=&e^{\imath\Phi(x,y)}\int\frac{d^4p}{(2\pi)^4}
e^{-\imath p\cdot(x-y)}e^{-p_t^2/h}\nonumber\\
&&\!\!\!\!\times\sum_{n=0}^{\infty}2(-1)^n
\left\{-\Si^+(\ov{p}_\mu\ga^{\mu}W_A+W_B)
L_{n-1}\left(\frac{2}{h}p_t^2\right)
+\Si^-(\ov{p}_\mu\ga^{\mu}W_C+W_D)L_{n}\left(\frac{2}{h}p_t^2\right)
\right.\nonumber\\&&\left.
+2L_{n-1}^1\left(\frac{2}{h}p_t^2\right)\vec{p}_t\cdot\vec{\ga}W_E
\right\}.\nonumber\\
\eea

Concentrating for now on the proper two-point function, $\G$, we make 
the following identifications for the summations involving the dressing 
functions
\bea
e^{-p_t^2/h}\sum_{n=0}^{\infty}2(-1)^{n-1}A(\ov{p},n)
L_{n-1}\left(\frac{2}{h}p_t^2\right)&=&\hat{A}(\ov{p}^2,p_t^2),
\nonumber\\
e^{-p_t^2/h}\sum_{n=0}^{\infty}2(-1)^{n-1}B(\ov{p},n)
L_{n-1}\left(\frac{2}{h}p_t^2\right)&=&\hat{B}(\ov{p}^2,p_t^2),
\nonumber\\
e^{-p_t^2/h}\sum_{n=0}^{\infty}2(-1)^{n}C(\ov{p},n)
L_{n}\left(\frac{2}{h}p_t^2\right)&=&\hat{C}(\ov{p}^2,p_t^2),
\nonumber\\
e^{-p_t^2/h}\sum_{n=0}^{\infty}2(-1)^{n}D(\ov{p},n)
L_{n}\left(\frac{2}{h}p_t^2\right)&=&\hat{D}(\ov{p}^2,p_t^2),
\nonumber\\
e^{-p_t^2/h}\sum_{n=0}^{\infty}4(-1)^{n-1}E(\ov{p},n)
L_{n-1}^{1}\left(\frac{2}{h}p_t^2\right)&=&\hat{E}(\ov{p}^2,p_t^2).
\label{eq:sumid0}
\eea
At tree-level,
\bea
\hat{A}^{(0)}(\ov{p}^2,p_t^2)=\hat{C}^{(0)}(\ov{p}^2,p_t^2)
=\hat{E}^{(0)}(\ov{p}^2,p_t^2)&=&1,\nonumber\\
\hat{B}^{(0)}(\ov{p}^2,p_t^2)=\hat{D}^{(0)}(\ov{p}^2,p_t^2)&=&m,
\eea
as before.  Further, when $h=0$, the functions should reduce to the 
Landau gauge case, i.e.,
\be
\hat{A}_{h=0}(\ov{p}^2,p_t^2)=\hat{C}_{h=0}(\ov{p}^2,p_t^2)
=\hat{E}_{h=0}(\ov{p}^2,p_t^2)=A_L(p),\;\;
\hat{B}_{h=0}(\ov{p}^2,p_t^2)=\hat{D}_{h=0}(\ov{p}^2,p_t^2)=B_L(p).
\label{eq:lredux0}
\ee
The proper two-point function thus reads
\be
-\imath\G(x,y)=e^{\imath\Phi(x,y)}\int\frac{d^4p}{(2\pi)^4}
e^{-\imath p\cdot(x-y)}
\left\{\Si^+(\ov{p}_\mu\ga^{\mu}\hat{A}-\hat{B})
+\Si^-(\ov{p}_\mu\ga^{\mu}\hat{C}-\hat{D})
-\vec{p}_t\cdot\vec{\ga}\hat{E}\right\},
\label{eq:invprop}
\ee
where the dressing functions are now functions of $\ov{p}^2$ and 
$p_t^2$ (implicitly, they are also dependent on $h$).  The dressing 
functions will be determined by the gap equation.

The situation for the propagator is somewhat less straightforward.  We 
notice that in \eq{eq:sumid0}, if we were to replace $A(\ov{p},n)$ with 
$\hat{A}(\ov{p}^2,p_t^2)$ (similarly for the functions $B$-$E$) before 
summing over $n$, the summations reduce to the earlier forms in 
\eq{eq:nsums0} such that the above result, \eq{eq:invprop}, and in 
particular its Dirac structure, would be unchanged.  Let us assume for 
now that this is also true for the summations involving the propagator, 
at least for small $h$ (we will justify this later).  In effect, the 
approximation retains the explicit factors of $n$ in the expressions 
and neglects any implicit functional dependence on $n$ in order to 
perform the summation.  Further, knowing that the functions should 
reduce to their Landau gauge counterparts in the limit $h\rightarrow0$, 
let us assume that we may expand around $h=0$.  Having set $F=0$, the 
denominator structure inherent to the propagator functions $W_A$-$W_E$, 
\eq{eq:propden}, reads
\be
\De=\De_1^2-\ov{p}^2\De_2^2,
\ee
with
\be
\De_1=\ov{p}^2\hat{A}\hat{C}-\hat{B}\hat{D}-2nh\hat{E}^2,\;\;
\De_2=\hat{A}\hat{D}-\hat{B}\hat{C}.
\ee
Under the above assumptions, the factor $\De_2$ will vanish as 
$h\rightarrow0$.  Expanding in $\De_2$, we then have that
\be
\frac{1}{\De}=\frac{1}{\De_1^2}+\co(\De_2^2).
\ee
To first order in $\De_2$, the propagator functions, \eq{eq:propfac} 
are now
\bea
W_A=\frac{\hat{C}}{\De_1}-\frac{\De_2\hat{D}}{\De_1^2},\;\;
W_B=\frac{\hat{D}}{\De_1}-\frac{\ov{p}^2\De_2\hat{C}}{\De_1^2},\;\;
W_C=\frac{\hat{A}}{\De_1}+\frac{\De_2\hat{B}}{\De_1^2},\;\;
W_D=\frac{\hat{B}}{\De_1}+\frac{\ov{p}^2\De_2\hat{A}}{\De_1^2},\;\;
W_E=\frac{\hat{E}}{\De_1}.
\eea
The Feynman prescription for the denominator factor may be assigned (so 
as to agree with the tree-level result) and we write,
\be
\De_1=\ov{p}^2\hat{A}\hat{C}-\hat{B}\hat{D}-2nh\hat{E}^2+\imath0_+
=-h\hat{E}^2[\ro+2n],\;\;\al=\frac{p_t^2}{h},
\ee
redefining $\ro$ from the earlier expression, \eq{eq:rodef0}.  The 
propagator now reads
\bea
\imath S(x,y)&=&e^{\imath\Phi(x,y)}\int\frac{d^4p}{(2\pi)^4}
e^{-\imath p\cdot(x-y)}\nonumber\\
&&\!\!\!\!\times\left\{e^{-\al}\sum_{n=0}^{\infty}
\frac{2(-1)^n}{(-h\hat{E}^2)[\ro+2n]}
\left[-\Si^+(\ov{p}_{\mu}\ga^{\mu}\hat{C}+\hat{D})L_{n-1}(2\al)
+\Si^-(\ov{p}_{\mu}\ga^{\mu}\hat{A}+\hat{B})L_{n}(2\al)
+2\vec{p}_t\cdot\vec{\ga}\hat{E}L_{n-1}^1(2\al)\right]
\right.\nonumber\\&&\left.
+e^{-\al}\sum_{n=0}^{\infty}
\frac{2(-1)^n(\hat{A}\hat{D}-\hat{B}\hat{C})}{(-h\hat{E}^2)^2[\ro+2n]^2}
\left[\Si^+(\ov{p}_{\mu}\ga^{\mu}\hat{D}
+\hat{p}^2\hat{C})L_{n-1}(2\al)
+\Si^-(\ov{p}_{\mu}\ga^{\mu}\hat{B}+\ov{p}^2\hat{A})L_{n}(2\al)\right]
\right\}.
\eea
Applying the results for the summations using \eq{eq:propint0} (for the 
double denominator factors, one must differentiate with respect to $\ro$) 
one arrives at the parametric form for the propagator
\bea
\imath S(x,y)&=&e^{\imath\Phi(x,y)}\int\frac{d^4p}{(2\pi)^4}
e^{-\imath p\cdot(x-y)}(-\imath)\int_0^{\infty}ds
\exp{\left\{\imath
\left[s\frac{[\ov{p}^2\hat{A}\hat{C}-\hat{B}\hat{D}+\imath0_+]}
{\hat{E}^2}-\frac{p_t^2}{h}\tan{(sh)}\right]\right\}}\nonumber\\
&&\times\left\{\frac{1}{\hat{E}^2}
\left([1-\imath\tan{(sh)}]\Si^+(\ov{p}_{\mu}\ga^{\mu}\hat{C}+\hat{D})
+[1+\imath\tan{(sh)}]\Si^-(\ov{p}_{\mu}\ga^{\mu}\hat{A}+\hat{B})
-[1+\tan^2{(sh)}]\vec{p}_t\cdot\vec{\ga}\hat{E}\right)
\right.\nonumber\\&&\left.
+\frac{(-\imath s)(\hat{A}\hat{D}-\hat{B}\hat{C})}{\hat{E}^4}
\left(-[1-\imath\tan{(sh)}]
\Si^+(\ov{p}_{\mu}\ga^{\mu}\hat{D}+\ov{p}^2\hat{C})
+[1+\imath\tan{(sh)}]
\Si^-(\ov{p}_{\mu}\ga^{\mu}\hat{B}+\ov{p}^2\hat{A})\right)\right\}.
\eea
In the absence of the coupling to the gluons, this expression explicitly 
reduces to the tree-level case from earlier, \eq{eq:ptree0}, for 
arbitrary magnetic field.  Expanding (to first order) in $h$ and 
evaluating the parametric integrals, we arrive at the following 
expression for the dressed quark propagator:
\bea
\imath S(x,y)&=&e^{\imath\Phi(x,y)}\int\frac{d^4p}{(2\pi)^4}
e^{-\imath p\cdot(x-y)}
\left\{
\frac{1}
{[\ov{p}^2\hat{A}\hat{C}-p_t^2\hat{E}^2-\hat{B}\hat{D}+\imath0_+]}
\left[\Si^+(\ov{p}_{\mu}\ga^{\mu}\hat{C}+\hat{D})
+\Si^-(\ov{p}_{\mu}\ga^{\mu}\hat{A}+\hat{B})
-\vec{p}_t\cdot\vec{\ga}\hat{E}\right]
\right.\nonumber\\&&\left.
+\frac{h\hat{E}^2}
{[\ov{p}^2\hat{A}\hat{C}-p_t^2\hat{E}^2-\hat{B}\hat{D}+\imath0_+]^2}
\left[\Si^+(\ov{p}_{\mu}\ga^{\mu}\hat{C}+\hat{D})
-\Si^-(\ov{p}_{\mu}\ga^{\mu}\hat{A}+\hat{B})\right]
\right.\nonumber\\&&\left.
+\frac{(\hat{A}\hat{D}-\hat{B}\hat{C})}
{[\ov{p}^2\hat{A}\hat{C}-p_t^2\hat{E}^2-\hat{B}\hat{D}+\imath0_+]^2}
\left[-\Si^+(\ov{p}_{\mu}\ga^{\mu}\hat{D}+\ov{p}^2\hat{C})
+\Si^-(\ov{p}_{\mu}\ga^{\mu}\hat{B}+\ov{p}^2\hat{A})\right]
\right\}.
\label{eq:prop}
\eea
One can see that as $h\rightarrow0$, the propagator reduces to its 
Landau gauge form if \eq{eq:lredux0} holds.  Further, both the 
nonperturbative propagator above, \eq{eq:prop}, and the proper two-point 
function, \eq{eq:invprop}, have the assumed form \eq{eq:anz0} such that 
the gap equation has the form \eq{eq:gap}.

The nature of the approximations used to derive \eq{eq:prop} are now 
clear.  By neglecting the $n$-dependence of the dressing functions and 
focusing on the small $h$ limit, we are able to perform the summation 
over the Landau levels.  The resulting expression reduces to both its 
tree-level form (in the absence of the gluon interaction) and its 
standard Landau gauge form (in the absence of the magnetic field).  The 
omission of the $n$-dependence within the summation is mitigated, 
because it is the gap equation (i.e., the dynamics) that will ultimately 
decide on the $p_t^2$-dependence of the dressing functions 
($\hat{A}$-$\hat{E}$) that replaces the $n$-dependence.  The Ritus 
decomposition was useful because it relates the Dirac structures of 
the proper two-point function to those of the propagator in the presence 
of the magnetic field -- the approximation is such that the connection 
between these Dirac structures is maintained, at least to leading order.  
Moreover, if one considers previous expressions, such as the denominator 
structure of the tree-level propagator under the Ritus decomposition, 
$(\ov{p}^2-2nh-m^2)$, one sees that for dimensional reasons, the 
$n$-dependence is accompanied by a factor of $h$, or is otherwise 
connected to the explicit indices of the Hermite or Laguerre polynomials 
(which are included in the summation).  Within the (dimensionless) 
dressing functions, this would presumably mean that for small $h$, the 
effect of any $n$-dependence is suppressed.

The expression for the chiral condensate, \eq{eq:cond0}, arising from 
the propagator, \eq{eq:prop}, is
\be
\ev{\ov{q}q}=-2\imath N_c\int\frac{d^4p}{(2\pi)^4}
\left\{\frac{\hat{B}+\hat{D}}
{[\ov{p}^2\hat{A}\hat{C}-p_t^2\hat{E}^2-\hat{B}\hat{D}+\imath0_+]}
+\frac{h\hat{E}^2(\hat{D}-\hat{B})
+\ov{p}^2(\hat{A}\hat{D}-\hat{B}\hat{C})
(\hat{A}-\hat{C})}
{[\ov{p}^2\hat{A}\hat{C}-p_t^2\hat{E}^2-\hat{B}\hat{D}+\imath0_+]^2}
\right\}.
\ee
Associated with the chiral condensate is the magnetic moment 
\cite{Frasca:2011zn,Buividovich:2009ih}, which we define as
\be
\ev{\ov{q}\Si^{12}q}=-N_c\mbox{Tr}_d\Si^{12}S(x,x),
\ee
where
\be
\Si^{12}=\frac{1}{2\imath}(\ga^1\ga^2-\ga^2\ga^1)=-(\Si^+-\Si^-).
\ee
Written in terms of $\Si^\pm$, it is clear that the magnetic moment is 
a measure of the asymmetry between the $\Si^+$- and $\Si^-$-projected 
Dirac components induced by the presence of the magnetic field.  As 
$h\rightarrow0$, the magnetic moment will vanish.  The explicit 
expression reads
\be
\ev{\ov{q}\Si^{12}q}=2\imath N_c\int\frac{d^4p}{(2\pi)^4}
\left\{\frac{\hat{B}-\hat{D}}
{[\ov{p}^2\hat{A}\hat{C}-p_t^2\hat{E}^2-\hat{B}\hat{D}+\imath0_+]}
-\frac{h\hat{E}^2(\hat{D}+\hat{B})
-\ov{p}^2(\hat{A}\hat{D}-\hat{B}\hat{C})
(\hat{A}+\hat{C})}
{[\ov{p}^2\hat{A}\hat{C}-p_t^2\hat{E}^2-\hat{B}\hat{D}+\imath0_+]^2}
\right\}
\label{eq:mom}
\ee
and one notices that this is equivalent to the expression for the 
condensate, up to certain minus sign factors.  The magnetic 
polarization, $\mu$, (which is a function of $h$) is the ratio of the 
magnetic moment to the condensate:
\be
\mu(h)=\left|\frac{\ev{\ov{q}\Si^{12}q}}{\ev{\ov{q}q}}\right|.
\label{eq:pol}
\ee
Further defining the function $\chi(h)$ with
\be
\chi(h)=\frac{\ev{\ov{q}\Si^{12}q}}{h\ev{\ov{q}q}},
\label{eq:chi}
\ee
the magnetic susceptibility is given by the limit
\be
\chi=\lim_{h\rightarrow0}\chi(h).
\ee

\section{Numerical results}
The gap equation, \eq{eq:gap}, is solved with the two phenomenological 
interactions, \eq{eq:int}, using the nonperturbative decompositions for 
the proper two-point function $\G$, \eq{eq:invprop}, and the propagator 
$S$, \eq{eq:prop}, via \eq{eq:anz0}.  We consider the chiral quark case, 
$m=0$.  As discussed, the Schwinger phase is an overall factor and can be 
dropped from the expressions.  After Wick rotating to Euclidean space 
($p_0\rightarrow\imath p_4$), the dressing functions are all functions 
of (longitudinal) $p_l^2=p_3^2+p_4^2$ and (transverse) 
$p_t^2=p_1^2+p_2^2$ momenta squared.  The gap equation is decomposed 
into a set of coupled scalar equations for the dressing functions 
(presented in Appendix~\ref{app:gapeqn} for completeness) which are 
solved iteratively using standard numerical techniques.  Notice that the 
exponential character of the phenomenological interaction means that all 
integrals are explicitly ultraviolet finite and there is no need for 
renormalization.

As mentioned previously, the type I interaction given in \eq{eq:int} 
corresponds to that used in a phenomenological study of dynamical chiral 
symmetry breaking and the light meson spectrum \cite{Alkofer:2002bp} 
(and is a simplified version of the interaction used in the earlier 
study of Ref.~\cite{Maris:1999nt}).  For the type I interaction, the 
parameter $d$ sets the overall amplitude of the interaction while $\w$ 
gives the position of its maximum.  In Ref.~\cite{Alkofer:2002bp}, a 
range of $\w$ values were considered and the values of $d$ and the quark 
masses fitted to reproduce the physical pseudoscalar meson masses and 
leptonic decay constants.  It was seen that once the parameter sets were 
fixed, the predicted vector, scalar, and $1^{++}$ axialvector meson 
masses were relatively stable for varying $\w$, with only the $1^{+-}$ 
axialvector channel showing significant $\w$-dependence (the $1^{+-}$ 
meson mass increasing with $\w$).  In this study, we consider various 
values of $\w$ and using the parameter set 
$\w=0.5\,\mbox{GeV}$, $d=16\,\mbox{GeV}^{-2}$ from 
Ref.~\cite{Alkofer:2002bp} as a basis, we then fix $d$ so as to 
approximately keep the quark chiral condensate fixed 
(for both interaction types).  This gives the parameter sets shown in 
Table~\ref{tab:int}.  It was verified that the numerical solutions for 
the equations in this study with $h=0$ match those for the equations in 
Ref.~\cite{Alkofer:2002bp} (for both interaction types).
\begin{table}[t]
\caption{\label{tab:int}Interaction parameters and results for the 
quark chiral condensate, $\ev{\ov{q}q}_{h=0}$ (evaluated in the absence 
of the magnetic field), and associated magnetic susceptibility $\chi$. 
 See text for details}
\begin{tabular}{|c|cc|ccc|}\hline
&$\w$&$d$&$\ev{\ov{q}q}_{h=0}$&$\chi$&$\chi\ev{\ov{q}q}_{h=0}$\\\hline
\multirow{4}{*}{type I\;\;}&$0.4\,\mbox{GeV}$&$48\,\mbox{GeV}^{-2}$
&$(-251\,\mbox{MeV})^3$&$-1.77\,\mbox{GeV}^{-2}$&$28.0\,\mbox{MeV}$\\
&$0.5\,\mbox{GeV}$&$16\,\mbox{GeV}^{-2}$
&$(-251\,\mbox{MeV})^3$&$-1.93\,\mbox{GeV}^{-2}$&$30.6\,\mbox{MeV}$\\
&$0.6\,\mbox{GeV}$&$7.6\,\mbox{GeV}^{-2}$
&$(-251\,\mbox{MeV})^3$&$-2.07\,\mbox{GeV}^{-2}$&$32.7\,\mbox{MeV}$\\
&$0.7\,\mbox{GeV}$&$4.67\,\mbox{GeV}^{-2}$
&$(-251\,\mbox{MeV})^3$&$-2.09\,\mbox{GeV}^{-2}$&$32.9\,\mbox{MeV}$\\
\hline
\multirow{3}{*}{type II\;\;}&$0.4\,\mbox{GeV}$&$119\,\mbox{GeV}^{-2}$
&$(-252\,\mbox{MeV})^3$&$-1.66\,\mbox{GeV}^{-2}$&$26.5\,\mbox{MeV}$\\
&$0.5\,\mbox{GeV}$&$41\,\mbox{GeV}^{-2}$
&$(-251\,\mbox{MeV})^3$&$-1.75\,\mbox{GeV}^{-2}$&$27.8\,\mbox{MeV}$\\
&$0.6\,\mbox{GeV}$&$17.4\,\mbox{GeV}^{-2}$
&$(-251\,\mbox{MeV})^3$&$-1.84\,\mbox{GeV}^{-2}$&$29.2\,\mbox{MeV}$\\
\hline
\end{tabular}
\end{table}

\begin{figure}[t]
\vspace{0.8cm}
\includegraphics[width=0.415\linewidth]{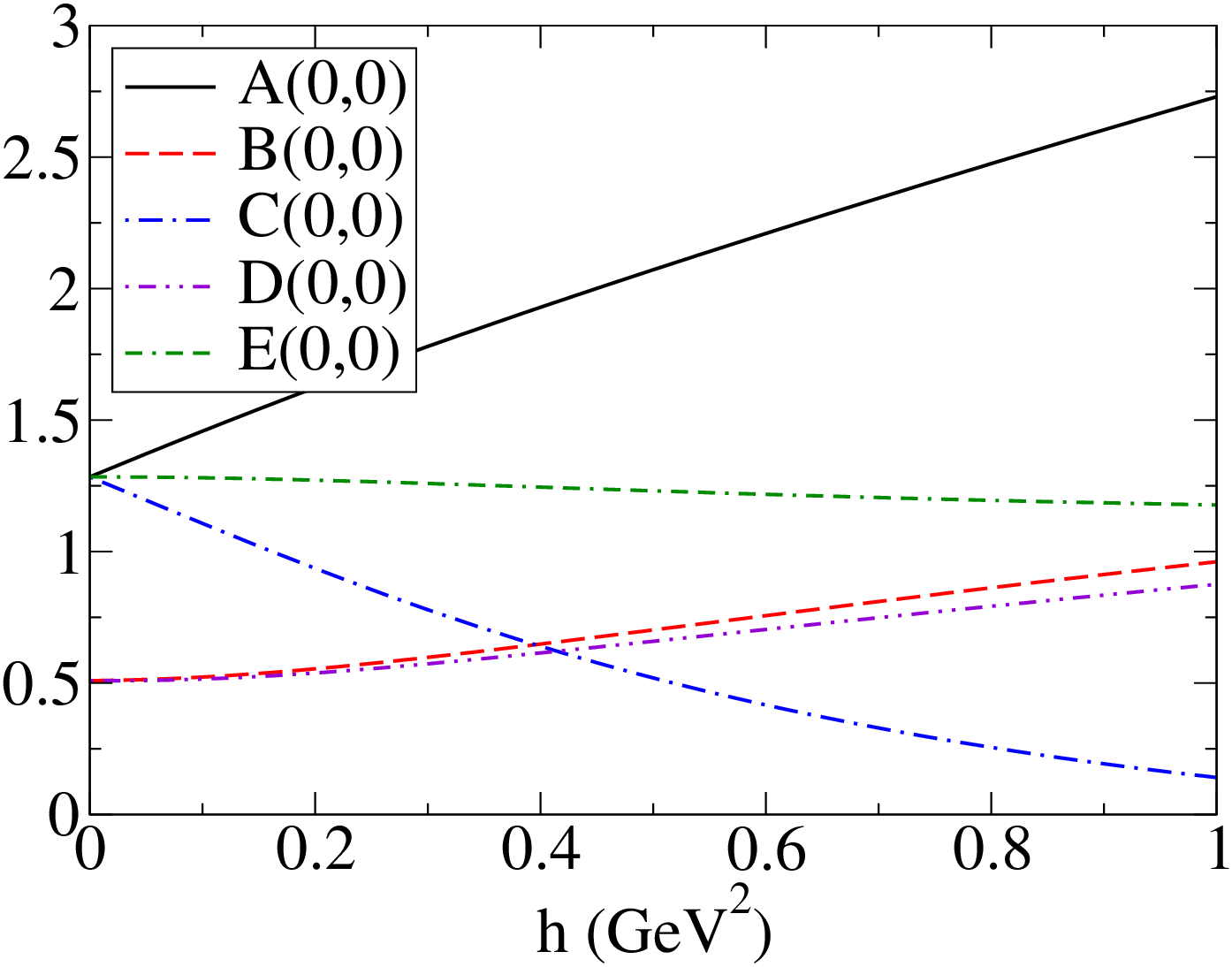}
\hspace{0.8cm}\includegraphics[width=0.4\linewidth]{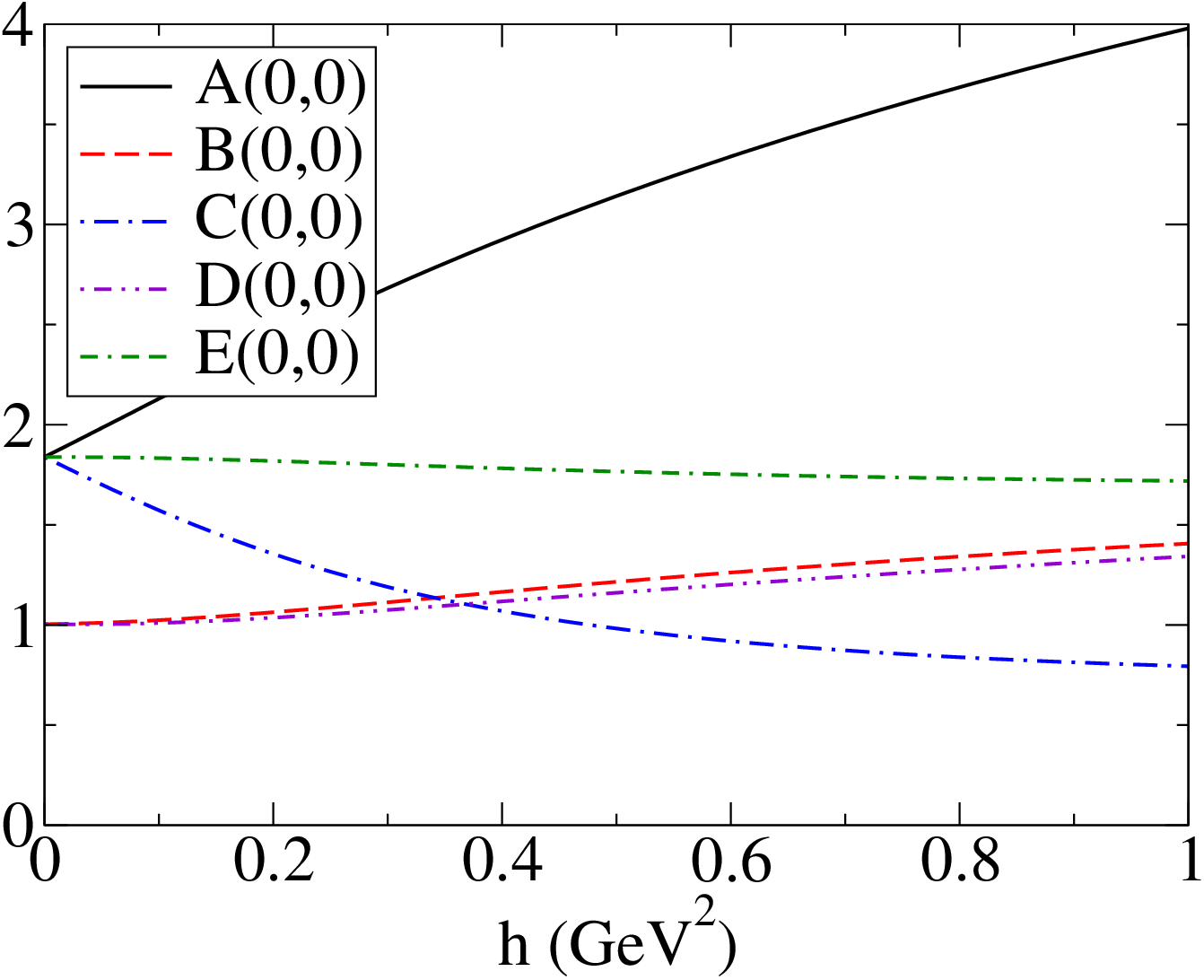}
\caption{\label{fig:dressh}Plot of the dressing functions 
$\hat{A}$-$\hat{E}$ evaluated at $p_l^2=p_t^2=0$ as a function of $h$ 
for the type I (left panel) and type II (right panel) interactions with 
$\w=0.5\,\mbox{GeV}$ used for both.  See text for details.}
\end{figure}

To illustrate the effect of the magnetic field on the propagator, the 
dressing functions $\hat{A}$-$\hat{E}$, evaluated at $p_l^2=p_t^2=0$ 
and for the $\w=0.5\,\mbox{GeV}$ parameter sets, are plotted as a 
function of $h$ for both types of interaction in Fig.~\ref{fig:dressh}.  
It is in the infrared that the effects of $h$ on the dressing functions 
are most prominent -- the dressing functions reduce to their tree-level 
forms in the ultraviolet.  The results for the different $\w$ parameter 
sets are similar.  One can see quite clearly, that as $h\rightarrow0$, 
the functions $\hat{A}$, $\hat{C}$, and $\hat{E}$ approach the same 
value, as do $\hat{B}$ and $\hat{D}$.  In the case of the type I 
interaction, the numerical values at $h=0$ can be explicitly compared 
with those presented in Ref.~\cite{Alkofer:2002bp}, showing that the 
standard Landau gauge results emerge as the magnetic field vanishes and 
\eq{eq:lredux0} is verified.  Comparing the two interactions, it is 
noticeable that the dressing functions show qualitatively the same 
pattern for varying values of $h$, although the numerical values for the 
dressing functions are rather different for the two types of interaction 
(the type I interaction vanishes, whereas the type II interaction is 
constant in the infrared).  This is to be expected, since the interaction 
is independent of the magnetic field under the truncation scheme 
considered in this study.

\begin{figure}[t]
\vspace{0.8cm}
\includegraphics[width=0.4\linewidth]{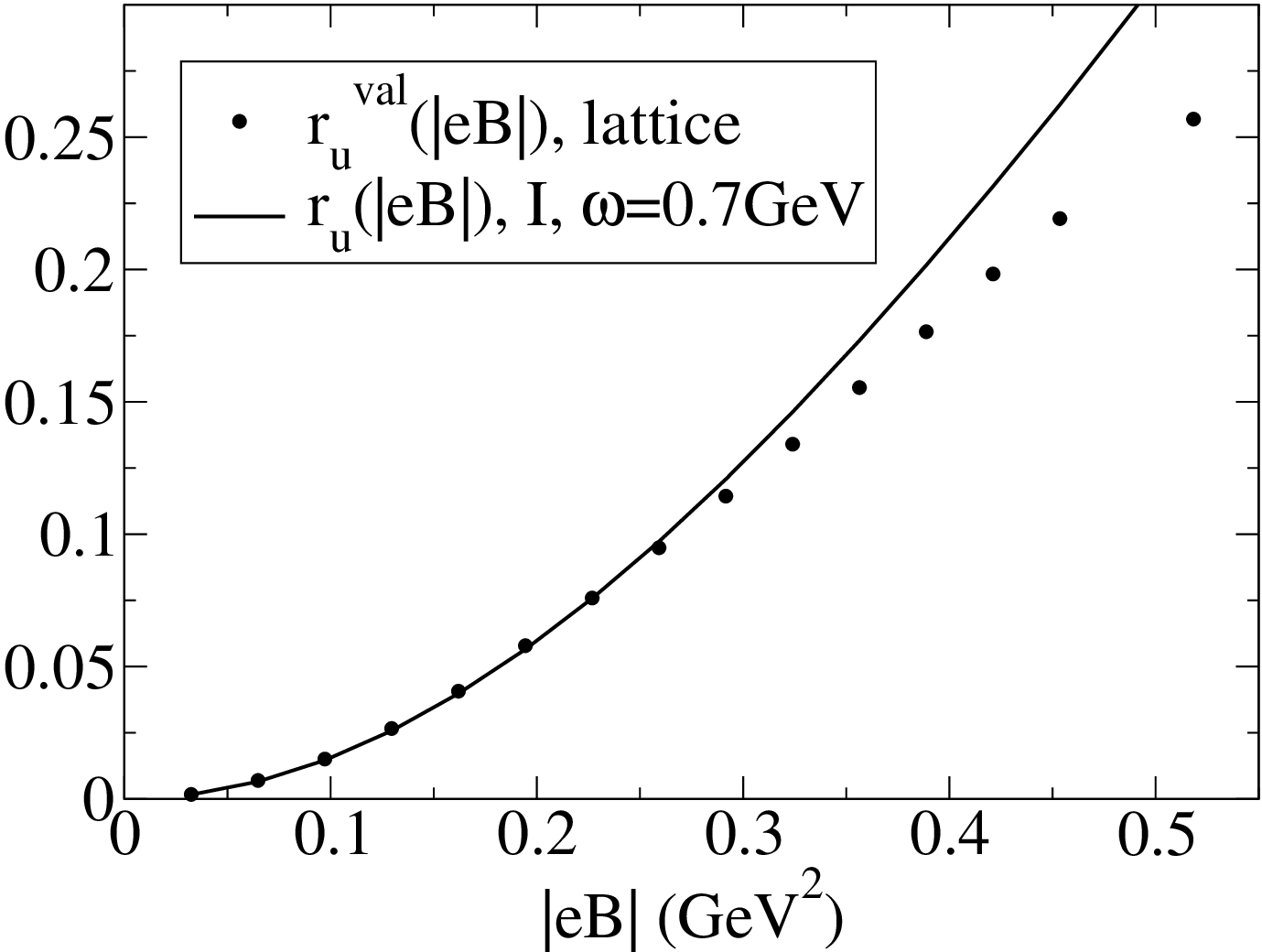}
\caption{\label{fig:rlat}Comparison of the up-quark relative increment 
(as a function of $|eB|$) for the type I, $\w=0.7\,\mbox{GeV}$ curve 
with the lattice results of Ref.~\cite{D'Elia:2011zu}.  
See text for details.}
\end{figure}

The chiral condensate in the presence of the magnetic field can be 
expressed in the form of the so-called relative increment, $r(h)$, 
defined via the dimensionless ratio
\be
r(h)=\frac{\ev{\ov{q}q}_h}{\ev{\ov{q}q}_{h=0}}-1.
\label{eq:rhdef}
\ee
\begin{figure}[t]
\vspace{0.8cm}
\includegraphics[width=0.4\linewidth]{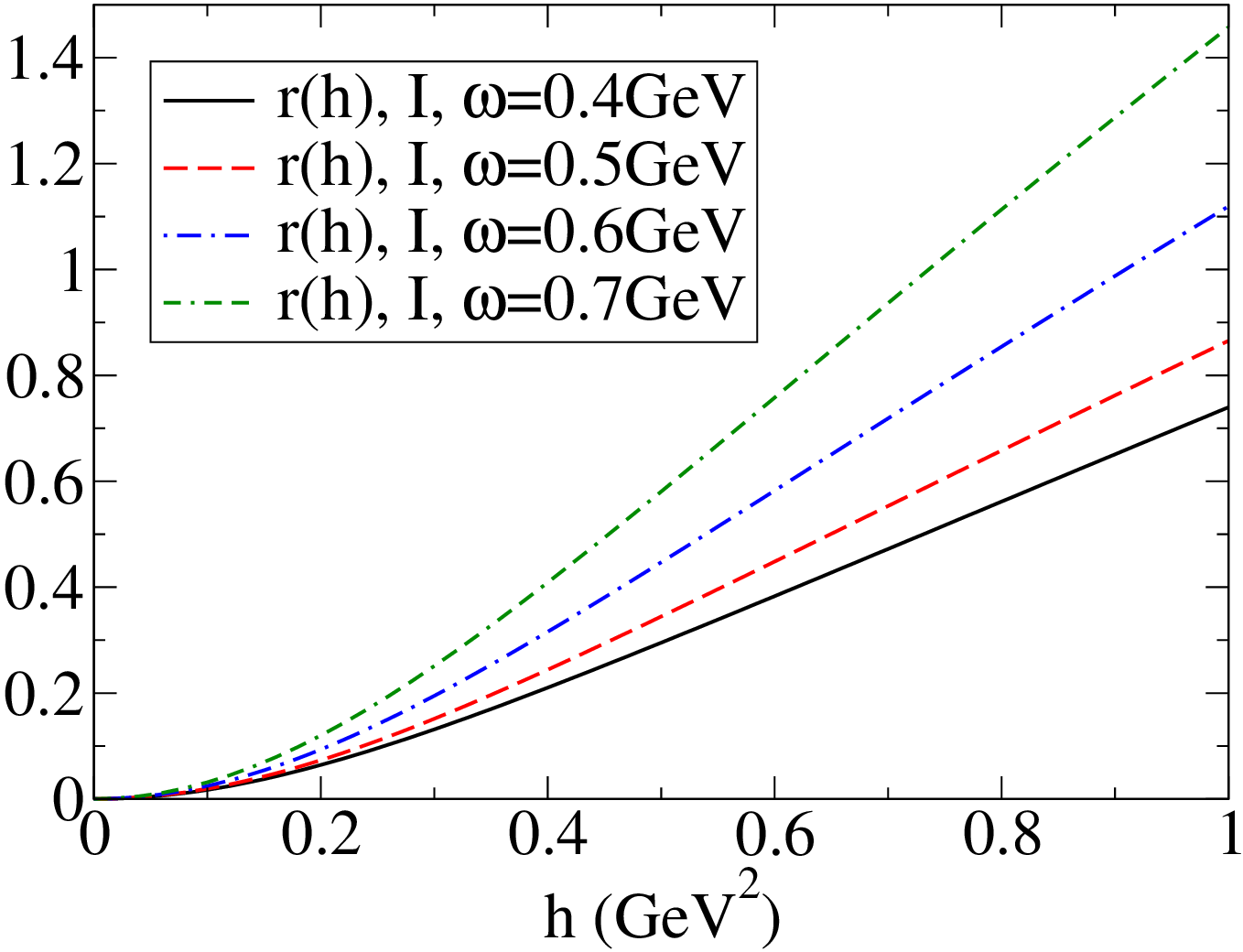}
\hspace{0.8cm}\includegraphics[width=0.4\linewidth]{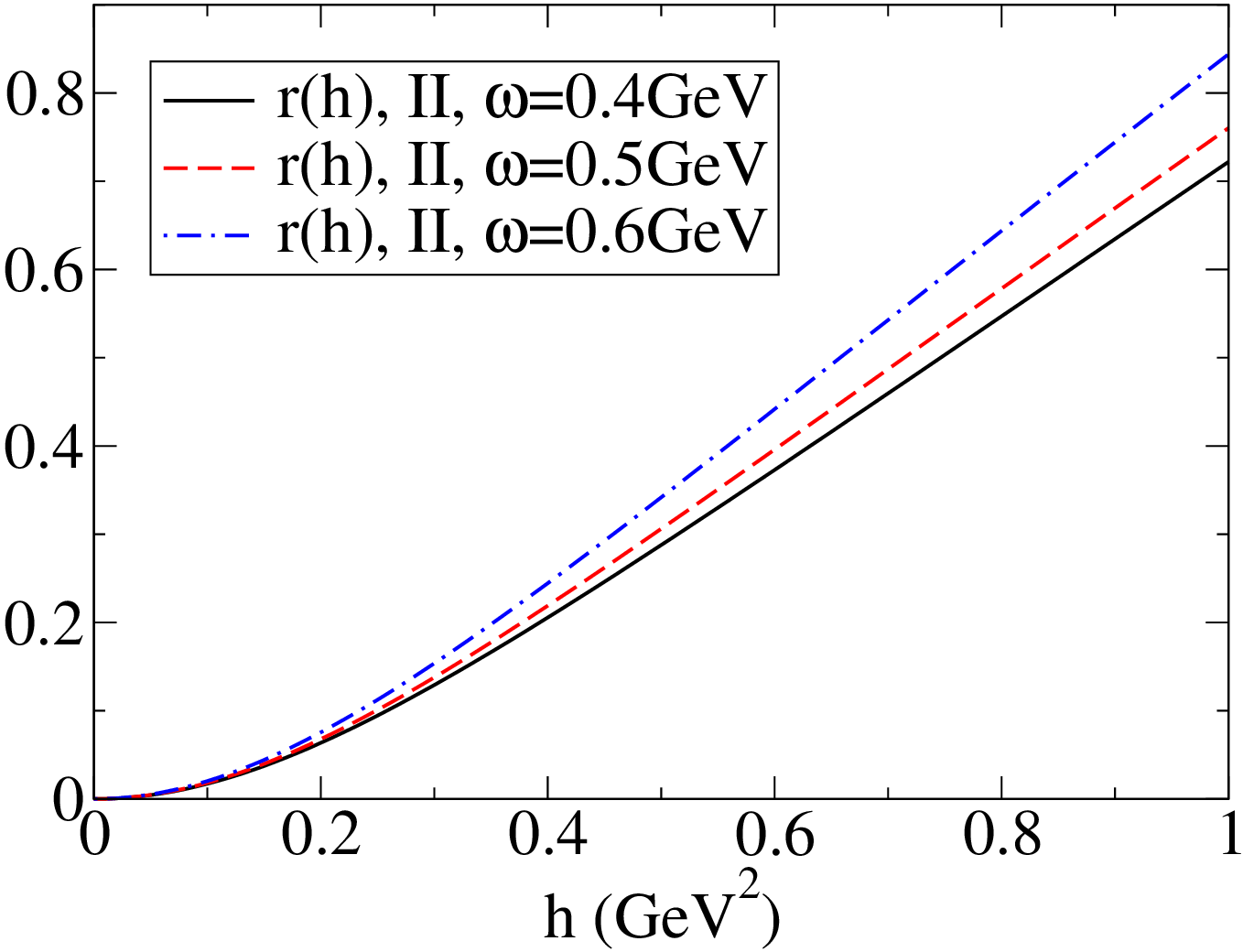}
\caption{\label{fig:rh}Plot of the (dimensionless) ratio $r(h)$, 
\eq{eq:rhdef}, as a function of $h$ for type I (left panel) and 
type II (right panel) interactions.  See text for details.}
\end{figure}
$r(h)$ is plotted for both types of interaction and varying $\w$ in 
Fig.~\ref{fig:rh}.  Recent lattice calculations, e.g., Refs.~\cite{D'Elia:2011zu,D'Elia:2013twa,Bali:2012zg,Bali:2013cf,Buividovich:2008wf}, 
indicate that for small values of $h$ ($<0.3\,\mbox{GeV}^2$) this ratio 
should rise quadratically with increasing $h$ and for large $h$, 
linearly.\footnote{We should point out that in contrast, the lattice 
study of Ref.~\cite{Ilgenfritz:2012fw} found that the small $h$ behavior 
is linear in the chiral limit, in accordance with the chiral 
perturbation theory result \cite{Shushpanov:1997sf}.}  This behavior is 
qualitatively well-reproduced by the results here.  We notice that $r(h)$ 
is larger for the type I interaction and increases with the 
parameter $\w$.  In the lattice calculation of Ref.~\cite{D'Elia:2011zu}, 
it was shown that unquenching effects are significant when considering 
the condensate in the presence of the magnetic field - the dynamical 
quark contribution was of the order of $40\%$.  Since the truncated gap 
equation studied here corresponds to a quenched system (although the 
parameter sets used stem from those fitted to physical observables in 
\cite{Alkofer:2002bp}), it makes sense to compare our results to those 
of Ref.~\cite{D'Elia:2011zu} that only include the valence quark 
contributions and we choose the up-quark for concreteness (this function 
is labelled $r_u^{\mbox{val}}$ in Ref.~\cite{D'Elia:2011zu}).  To make 
the comparison, we have to convert the argument $h=QB$ ($>0$) with the 
factor $Q=2e/3$ for the up-quark electric charge.  This is most easily 
done by fitting the curves for $r(h)$ with the formula 
(suggested in Ref.~\cite{D'Elia:2011zu})
\be
r(h)=a_0h\,\mbox{arctan}(a_1h).
\ee
The up-quark relative increment function is then given by
\be
r_u(|eB|)
=\frac{2}{3}a_0|eB|\,\mbox{arctan}\left(\frac{2}{3}a_1|eB|\right).
\ee
It is found that the type I interaction with $\w=0.7\,\mbox{GeV}$ 
(for which $a_0\approx1.18\,\mbox{GeV}^{-2}$ and 
$a_1\approx2.99\,\mbox{GeV}^{-2}$) best compares to the lattice 
results.  The comparison is shown in Fig.~\ref{fig:rlat} 
(the data for $r_u^{\mbox{val}}$ are extracted from Table I of 
Ref.~\cite{D'Elia:2011zu}  using the value $|eB|=b(180\,\mbox{MeV})^2$ 
to convert to physical units).  For small $|eB|$, the agreement is 
rather striking, whereas for large $|eB|$, the linear rise in $r_u$ 
has a larger coefficient than for the lattice result.  (Notice that in 
the lattice calculation of Ref.~\cite{D'Elia:2011zu}, the finite 
lattice spacing leads to saturation effects for very large magnetic 
fields and these tend to suppress $r_u^{\mbox{val}}$ in this region, as 
shown in Ref.~\cite{D'Elia:2013twa}.)  Recalling that the approximations 
used to derive the quark propagator in the presence of the magnetic 
field, \eq{eq:prop}, were tailored to the $h\rightarrow0$ limit, it is 
tempting to conclude from Fig.~\ref{fig:rlat} that the type I, 
$\w=0.7\,\mbox{GeV}$ curve is the preferred parametrization of the 
model interaction.  However, we urge some caution here.  In this study, 
the chiral quark condensate is considered whereas the lattice simulation 
of Ref.~\cite{D'Elia:2011zu} has finite bare quark masses corresponding 
to $m_\pi\approx200\,\mbox{MeV}$ and it may be that the direct 
comparison is not appropriate without taking this into account.  
Additionally, the various parameter sets used in this study are chosen 
so as to keep the condensate (in the absence of the magnetic field) 
fixed, using the type I, $\w=0.5\,\mbox{GeV}$ interaction from 
Ref.~\cite{Alkofer:2002bp} as a basis.  If one were to, for example, 
fix the parameters via a recalculation of the meson masses and leptonic 
decay constants, the parameter sets may change such that a considerably 
more detailed analysis would be required.  What is clear from 
Fig.~\ref{fig:rlat} though, is that the small $h$ quadratic behavior and 
the scale of the transition between the quadratic and linear regimes is 
well-reproduced.  It is worth pointing out that the upper estimate for 
the magnitude of magnetic fields in noncentral heavy-ion collisions 
\cite{Skokov:2009qp} is $|eB|\sim0.3\,\mbox{GeV}^2$, which 
coincidentally lies roughly at the transition scale between the small 
and large $h$ behaviors of the curves in Fig.~\ref{fig:rlat}.

\begin{figure}[t]
\vspace{0.8cm}
\includegraphics[width=0.4\linewidth]{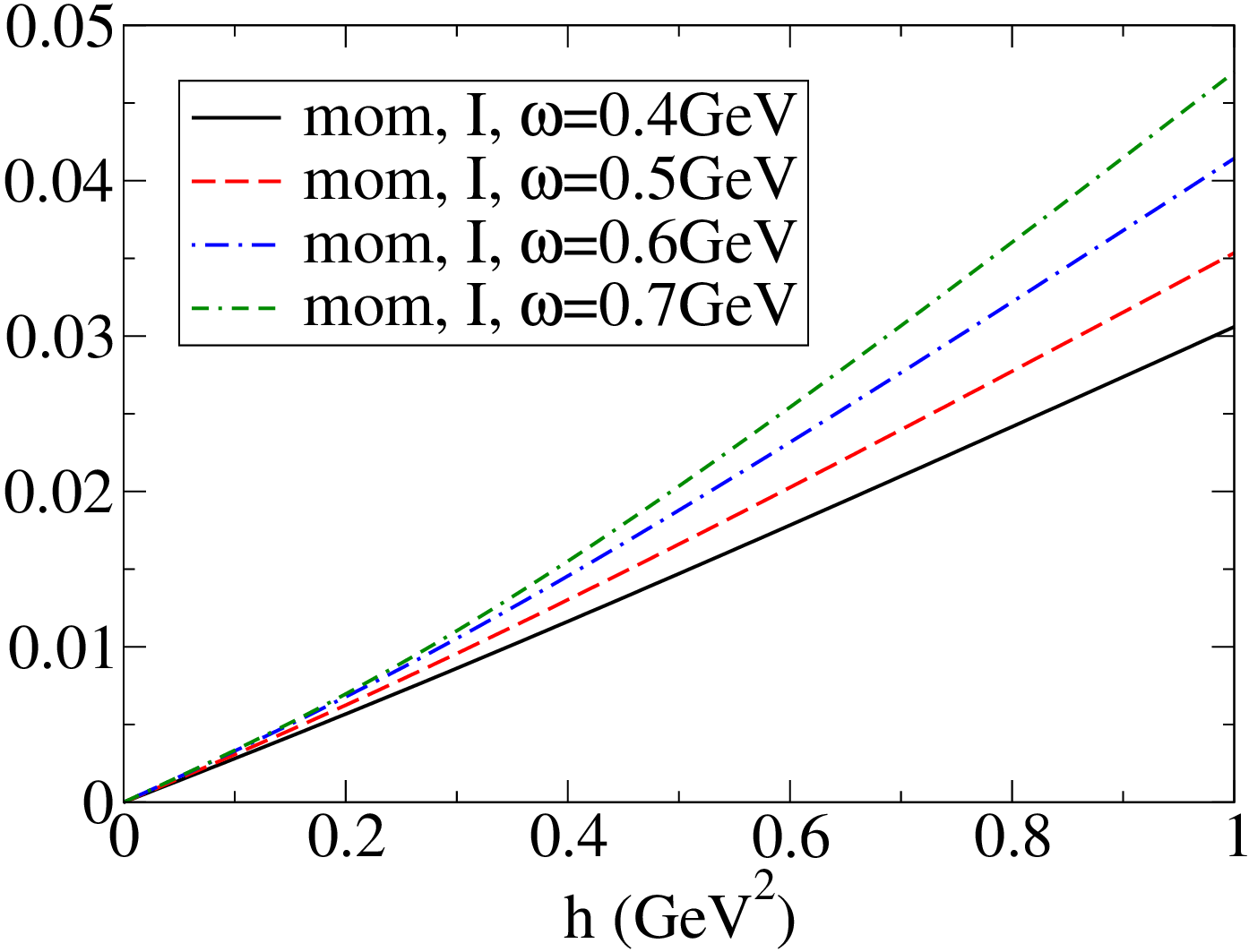}
\hspace{0.8cm}\includegraphics[width=0.4\linewidth]{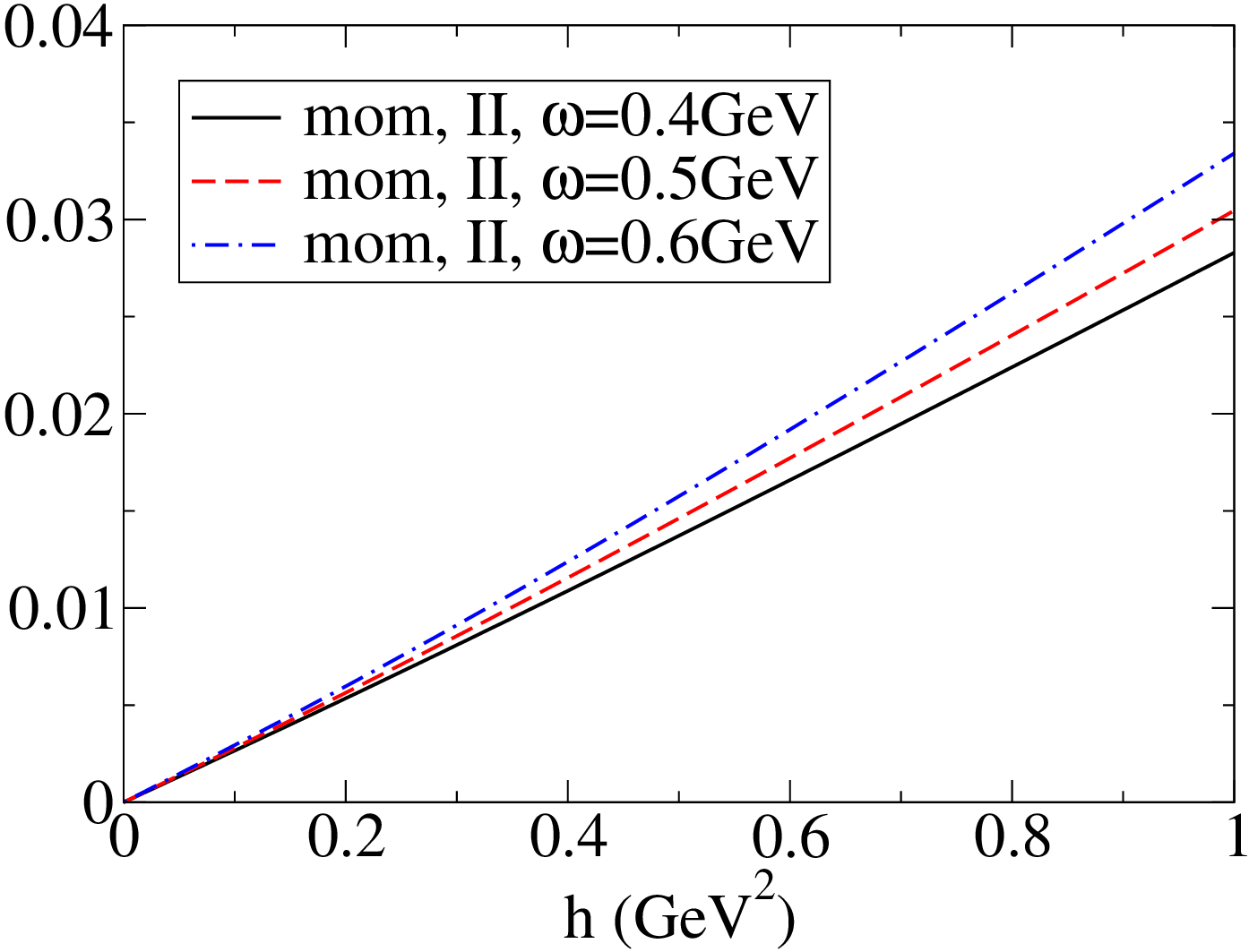}
\caption{\label{fig:mom}Plot of the magnetic moment 
$\ev{\ov{q}\Si^{12}q}$ (``mom'', in units of $\mbox{GeV}^{3}$) for the 
type I (left panel) and type II (right panel) interactions.  
See text for details.}
\end{figure}
\begin{figure}[t]
\vspace{0.8cm}
\includegraphics[width=0.4\linewidth]{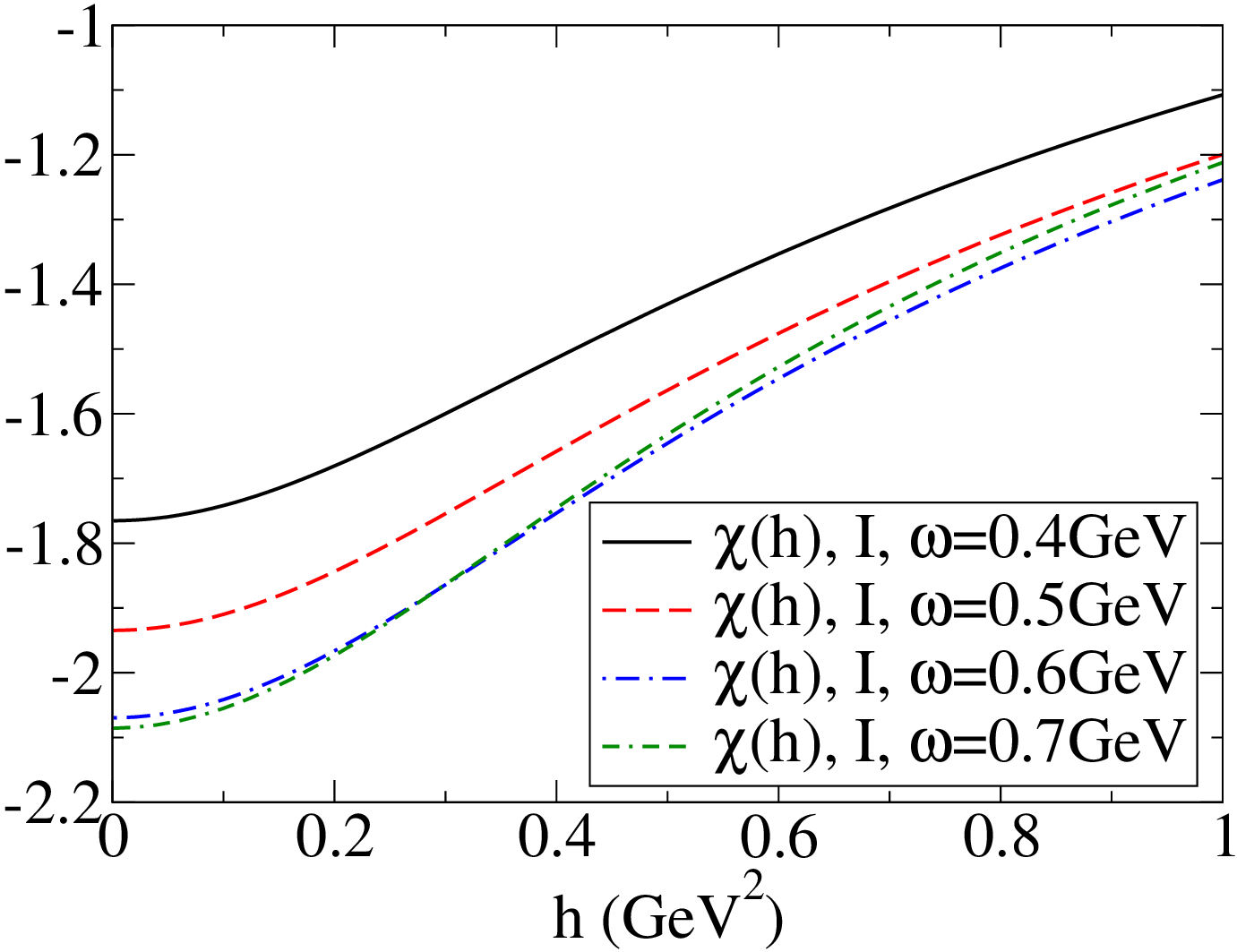}
\hspace{0.8cm}\includegraphics[width=0.4\linewidth]{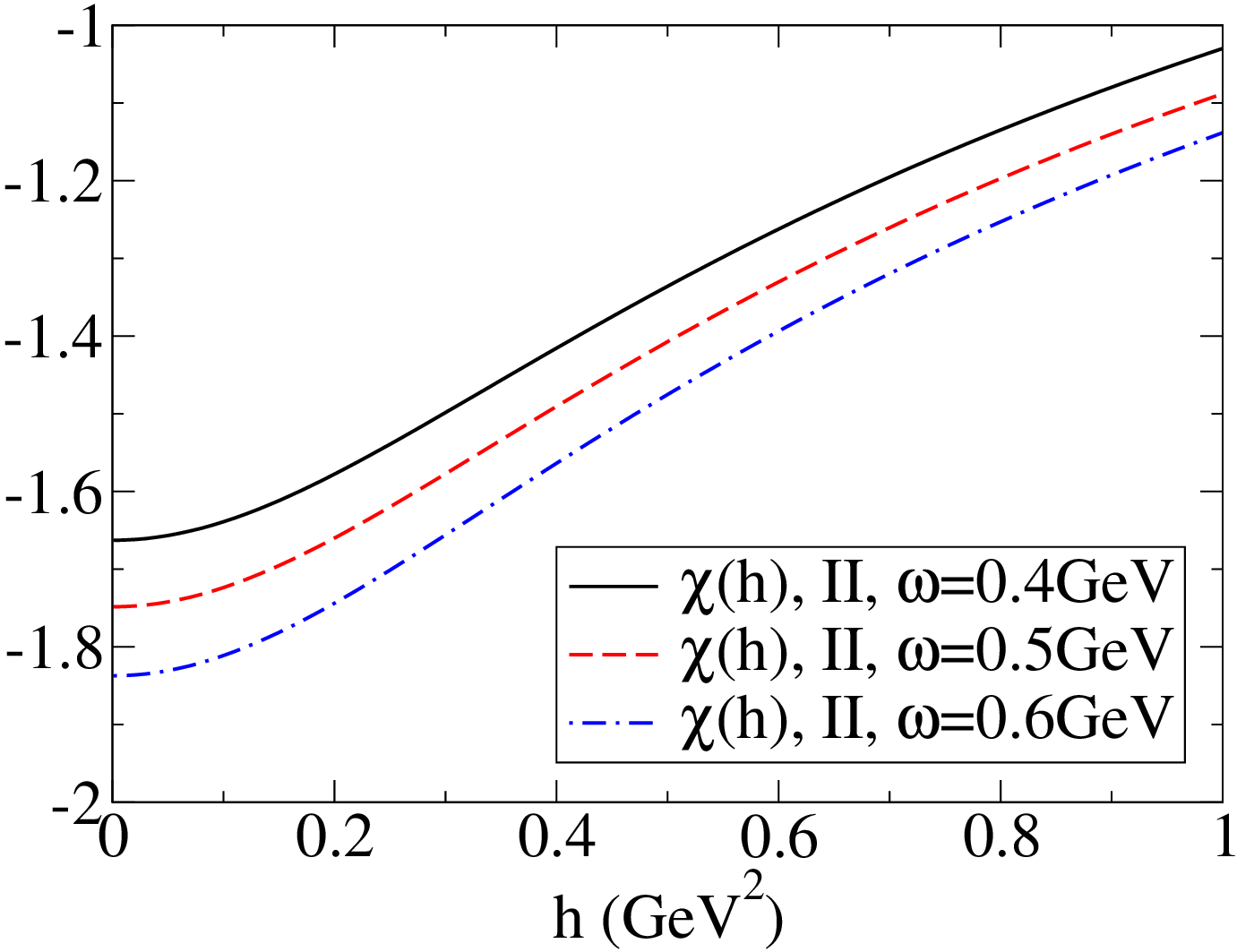}
\caption{\label{fig:chi}Plot of $\chi(h)$ (units of $\mbox{GeV}^{-2}$) 
for the type I (left panel) and type II (right panel) interactions.  
See text for details.}
\end{figure}
\begin{figure}[t]
\vspace{0.8cm}
\includegraphics[width=0.4\linewidth]{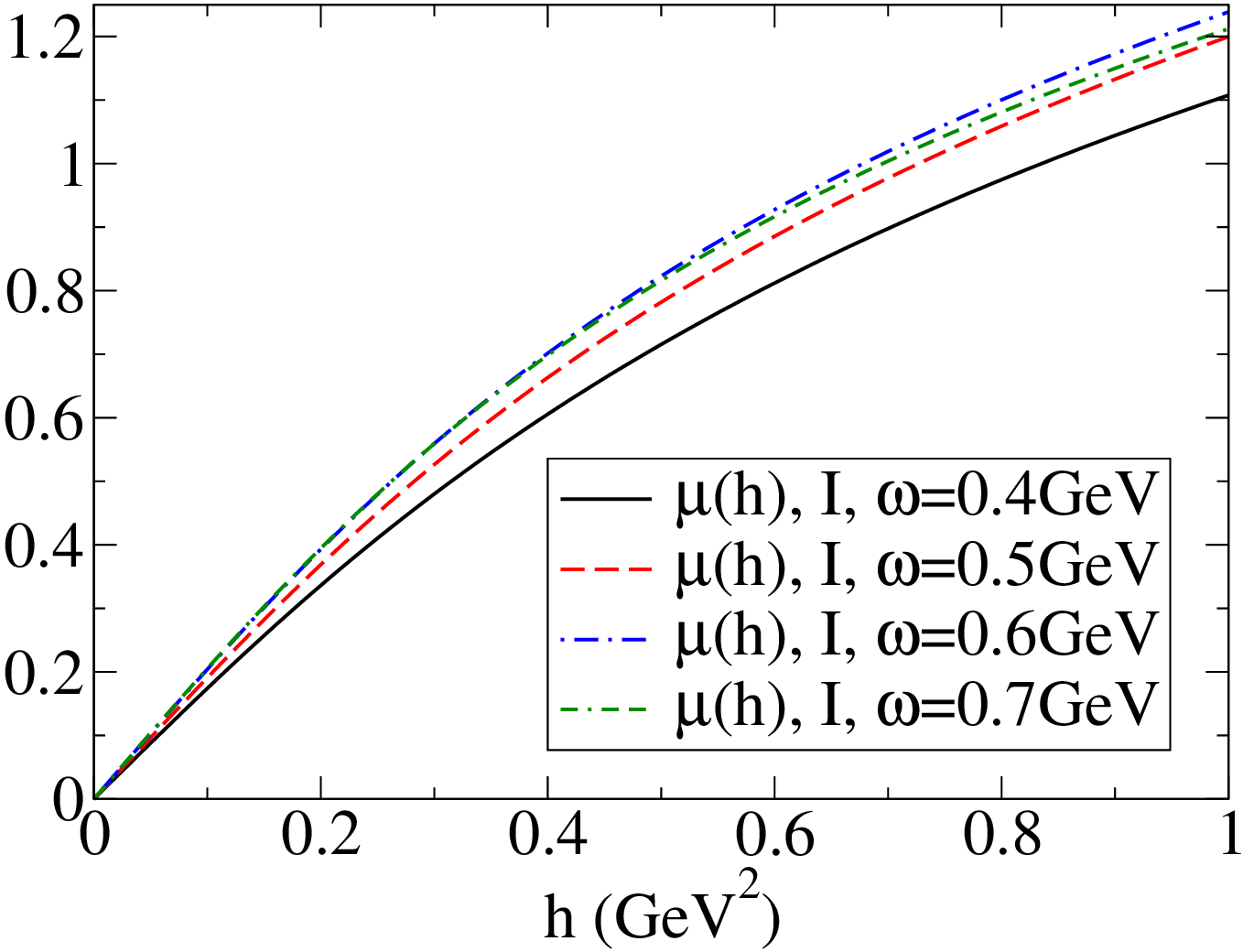}
\hspace{0.8cm}\includegraphics[width=0.4\linewidth]{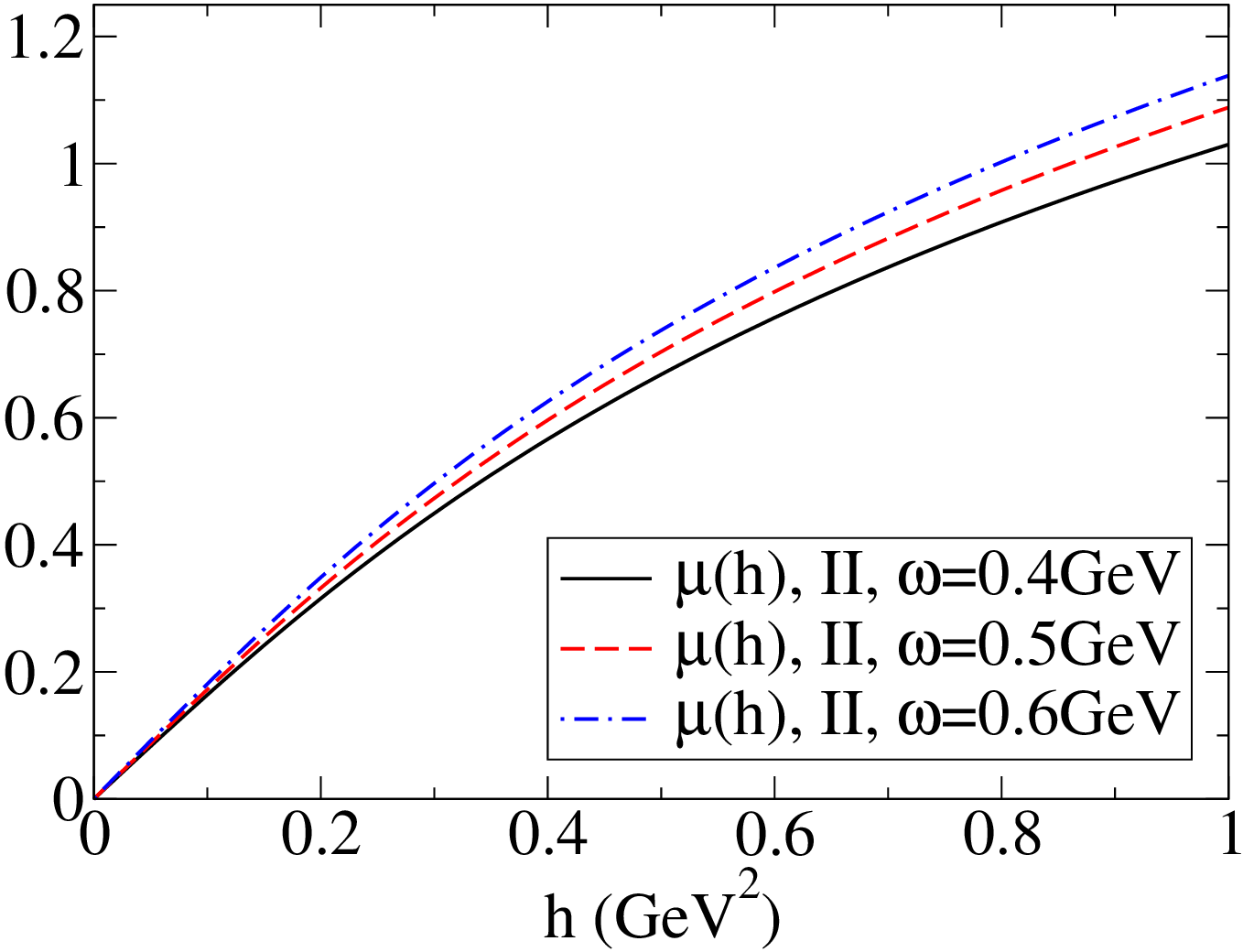}
\caption{\label{fig:mui}Plot of the magnetic polarization, $\mu(h)$, 
for the type I (left panel) and type II (right panel) interactions.  
See text for details.}
\end{figure}

The magnetic moment, $\ev{\ov{q}\Si^{12}q}$, \eq{eq:mom}, is plotted as 
a function of $h$ in Fig.~\ref{fig:mom}.  It is seen that 
$\ev{\ov{q}\Si^{12}q}$ is approximately linear for small $h$ 
(and vanishes for $h=0$, as it should) and that it follows much the same 
pattern of parameter dependence as for $r(h)$ in Fig.~\ref{fig:rh}.  
The related function $\chi(h)$, \eq{eq:chi}, is plotted in 
Fig.~\ref{fig:chi}.  The interesting part of this plot is the limit 
$h\rightarrow0$, from which the magnetic susceptibility, $\chi$, can 
be extracted.  It is seen that $\chi(h)$ approaches a constant for small 
values of $h$, the lowest values calculated 
(numerically $h=0.001\,\mbox{GeV}^2$ is the lowest nonzero value of $h$ 
considered) approximating $\chi$ and presented in Table~\ref{tab:int}.  
It is apparent that trying to extrapolate $\chi$ from $\chi(h)$ with, 
for example, only data for $h\geq0.2\,\mbox{GeV}^2$, the resulting 
values for $\chi$ would be rather inaccurate because of the bend in the 
curves around $h\sim0.1\,\mbox{GeV}^2$, underscoring the need for 
studying the small $h$ behavior emphasized in this study.  One other 
noticeable feature of Fig.~\ref{fig:chi} is that the type I interaction, 
$\w=0.7\,\mbox{GeV}$ curve appears outside the pattern of parameter 
dependence that one might have expected from the other curves.  
The explanation for this is quite straightforward.  Both the magnetic 
moment, $\ev{\ov{q}\Si^{12}q}$, plotted in Fig.~\ref{fig:mom} and the 
condensate, via the relative increment function, $r(h)$, plotted in 
Fig.~\ref{fig:rh} have a clear pattern of parameter dependence.  
However their ratio, the absolute value of which is the magnetic 
polarization, $\mu(h)$ given by \eq{eq:pol} and plotted in 
Fig.~\ref{fig:mui}, shows evidence for a turning point in $\mu(h)$ for 
increasing $\w$ (at least for the type I interaction).  Given that 
$|\chi(h)|=\mu(h)/h$, the $\w=0.7\,\mbox{GeV}$ curve seen from this 
perspective no longer appears out of place.

The results for the magnetic susceptibility, $\chi$, given in 
Table~\ref{tab:int} can be compared with those presented in 
Refs.~\cite{Frasca:2011zn,Buividovich:2009ih,Bali:2013cf}.  
The (quenched, chiral limit) lattice calculation of 
Ref.~\cite{Buividovich:2009ih} gives $\chi=-1.547(6)\,\mbox{GeV}^{-2}$, 
the (unquenched, finite bare quark mass) lattice results presented in 
Ref.~\cite{Bali:2013cf} have $\chi=-(2.08\pm0.08)\,\mbox{GeV}^{-2}$ 
(for the up-quark), whereas the calculations carried out in 
Ref.~\cite{Frasca:2011zn} give $\chi=-4.3\,\mbox{GeV}^{-2}$ 
(Nambu-Jona-Lasinio model) and $\chi=-5.25\,\mbox{GeV}^{-2}$ 
(quark-meson model).  Our results with 
$\chi\approx-(1.7-2.1)\,\mbox{GeV}^{-2}$ lie roughly between the 
lattice results, but are not inconsistent with the other calculations.  
Related to the susceptibility, $\chi$, the product 
$\chi\ev{\ov{q}q}_{h=0}$ is also of phenomenological interest.  
In Ref.~\cite{Frasca:2011zn}, various estimates for this quantity were 
reviewed, with the numerical values 
$\chi\ev{\ov{q}q}_{h=0}=40-70\,\mbox{MeV}$.  Our results, shown in 
Table~\ref{tab:int} lie in the range 
$\chi\ev{\ov{q}q}_{h=0}\approx28-33\,\mbox{MeV}$ and are somewhat 
smaller than this, although not dramatically so.

\begin{figure}[t]
\vspace{0.8cm}
\includegraphics[width=0.385\linewidth]{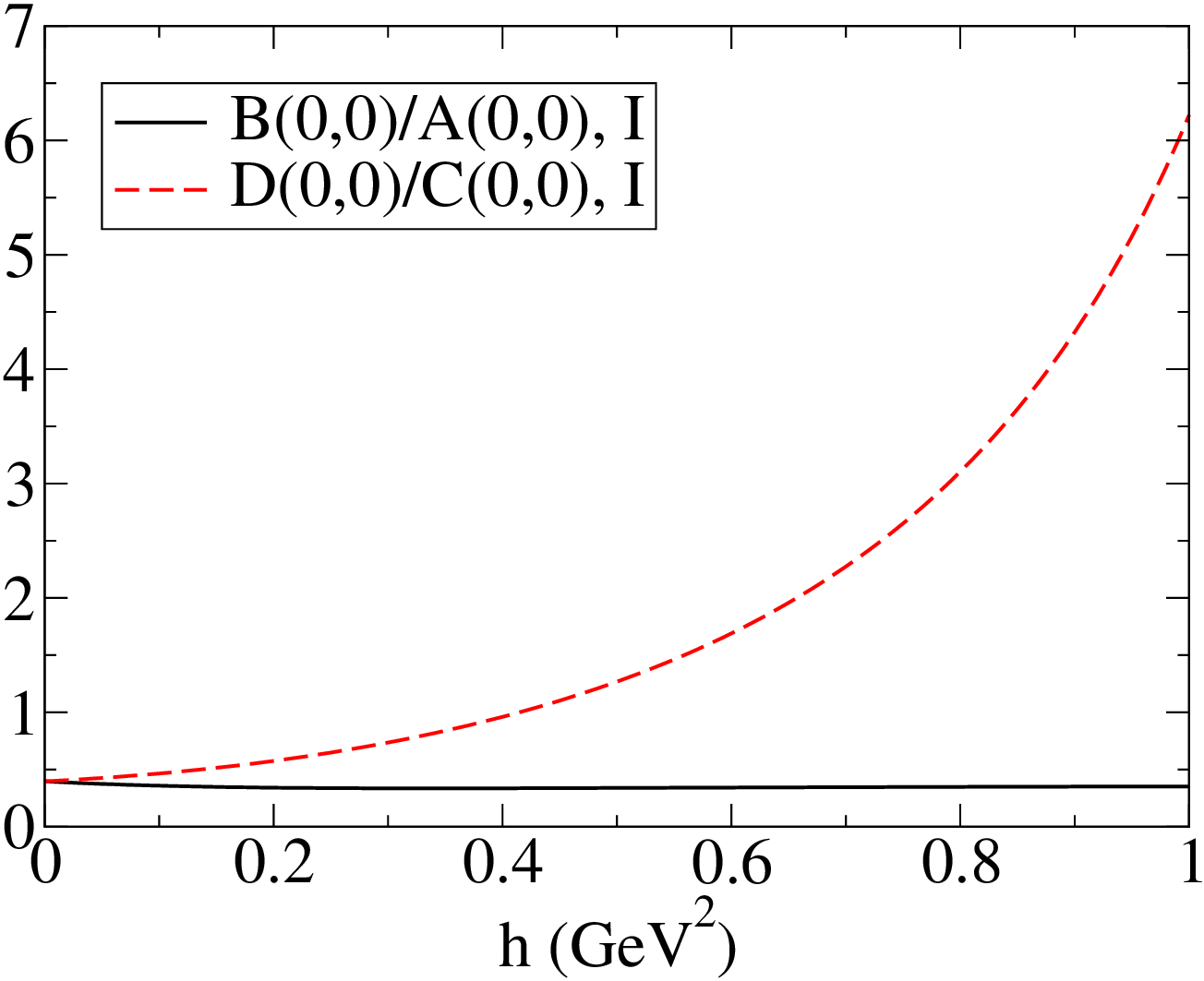}
\hspace{0.8cm}\includegraphics[width=0.4\linewidth]{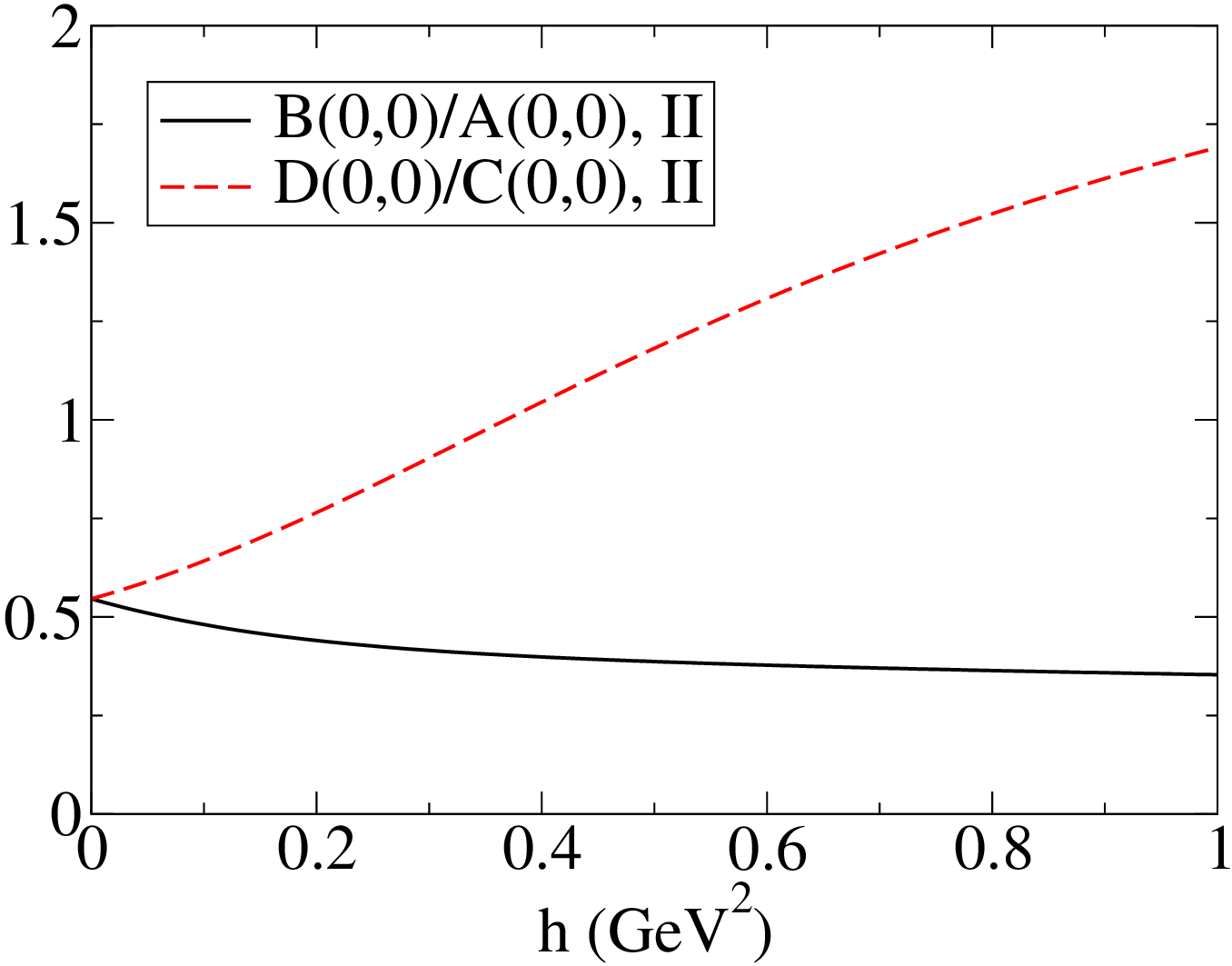}
\caption{\label{fig:dcba}Plot of ratios of dressing functions 
$\hat{B}/\hat{A}$ and $\hat{D}/\hat{C}$ (units of $\mbox{GeV}$) 
evaluated at $p_l^2=p_t^2=0$ as 
a function of $h$ for the type I (left panel) and type II (right panel) 
interactions, with $\w=0.5\,\mbox{GeV}$.  See text for details.}
\end{figure}

The magnetic polarization $\mu(h)$, defined in \eq{eq:pol}, is plotted 
in Fig.~\ref{fig:mui}.  In the lattice study of 
Ref.~\cite{Buividovich:2009ih}, it was shown that for large $h$, there 
is a saturation and $\mu(h)\rightarrow1$ as $h\rightarrow\infty$.  
The explanation for this is based on the expected dominance of the 
lowest Landau level for large magnetic fields.  For the lowest Landau 
level, the propagator is projected with $\Si^-$ (as discussed earlier) 
and one has the analytic result $\ev{\ov{q}q}=\ev{\ov{q}\Si^{12}q}$.  
This behavior is not reproduced here and the reason is clear -- the 
approximations made in order to sum the Landau levels of the 
nonperturbative propagator were tailored to the opposite limit 
($h\rightarrow0$) and are not suitable for determining the large 
$h$ behavior.  (Recall that this was also seen for the comparison of 
$r(h)$ to the lattice data in Fig.~\ref{fig:rlat}.)
Interestingly though, we do see indirect evidence for the dominance of 
the lowest Landau level.  Having performed a summation in this study, 
there is no direct access to the lowest Landau level contributions.  
However, we can look at the different spin projected dressing functions, 
noting that only $\hat{C}$ and $\hat{D}$ contribute to the lowest 
Landau level.  Now, the dressing functions do not individually contain 
much direct information: for example, the dressing functions have 
different numerical values but the same condensate when comparing the 
two interaction types and various parameter sets of this study.  
Combinations of the dressing functions though, e.g., the condensate or 
the mass function, do form physically meaningful quantities which one 
may compare.  Let us thus consider the mass function that would arise 
from the lowest Landau level spin structure (the self-energy terms 
proportional to $\Si^-$), the ratio $\hat{D}/\hat{C}$ evaluated at 
$p_l^2=p_t^2=0$, and compare it to the ratio $\hat{B}/\hat{A}$ 
(associated with the $\Si^+$ component).  This is plotted in 
Fig.~\ref{fig:dcba} for both types of interaction and with 
$\w=0.5\,\mbox{GeV}$.  It is seen that the ratio $\hat{B}/\hat{A}$ is 
approximately constant with varying $h$ for both types of interaction.  
However, the ratio $\hat{D}/\hat{C}$ increases significantly with $h$ 
(dramatically so for the type I interaction).  The mechanism for this is 
the decrease of $\hat{C}$ in the infrared with increasing $h$ 
(see also Fig.~\ref{fig:dressh}).  The decrease in $\hat{C}$ is already 
present for small $h$ (where the approximations should be valid) and so 
by extrapolation, one would expect the $\Si^-$-projected mass function 
(i.e., that connected with the spin structure of the lowest Landau level) 
to be dominant at large $h$.

\section{Summary and conclusions}
In this paper, the quark gap equation in the presence of a constant 
external magnetic field was studied.  The rainbow truncation was 
employed, with two versions of a simple phenomenological one-gluon 
exchange interaction.  The focus was on the small magnetic field limit, 
motivated by three concerns: the theoretical desire to connect the 
results to standard results in the absence of the magnetic field, the 
fact that the strong interaction is typically stronger than any other 
interaction, and the wish to calculate the magnetic susceptibility.  In 
order to do this, a nonperturbative approximation to the quark propagator 
in the presence of a small magnetic field was constructed, utilizing 
results \cite{Gorbar:2013upa,Gorbar:2013uga} for the summation over the 
Landau levels that arise when considering the Ritus eigenfunction method 
\cite{Ritus:1978cj} as applied to the fermion gap equation 
\cite{Leung:1995mh,Lee:1997zj,Leung:2005xz,Ayala:2006sv,Rojas:2008sg}.

With the phenomenological interactions considered, it was found that the 
chiral condensate rises quadratically for small magnetic fields and 
linearly for large fields, in qualitative agreement with recent lattice 
results 
\cite
{D'Elia:2011zu,D'Elia:2013twa,Bali:2012zg,Bali:2013cf,Buividovich:2008wf}.  
Comparing to the (valence quark contribution to the) quark condensate 
lattice results of Ref.~\cite{D'Elia:2011zu}, it was seen that the 
position of the transition between small and large magnetic fields can be 
quantitatively reproduced (and gives an estimate for the range of 
validity of our approximation).  The delineation between the small and 
large magnetic field behavior lies roughly at the upper estimate for the 
magnitudes of fields present in noncentral heavy-ion collisions 
\cite{Skokov:2009qp}.  The calculated magnetic susceptibility also 
qualitatively agreed with recent results 
\cite{Frasca:2011zn,Buividovich:2009ih,Bali:2013cf}.  It was observed 
that the approximated propagator does not reproduce the expected large 
magnetic field behavior of the magnetic polarization, although this was 
unsurprising given the nature of the approximations tailored to the small 
field limit.

At a more technical level, it was seen that the response of the system to 
the magnetic field was dependent on both the form of the interaction and 
its parametrization, despite the fact that the chiral condensate in the 
absence of the magnetic field was held fixed.  This is in marked contrast 
to previous studies of the light meson mass spectrum and leptonic decay 
constants that employed this type of interaction 
\cite{Alkofer:2002bp,Maris:1999nt}, where it was observed that the 
results (aside from the $1^{+-}$ axialvector meson masses) were stable 
with respect to such parameter changes.  The conclusion is that the 
magnetic field is a sensitive probe of the quark self-energy and the 
details of the interaction.  From the phenomenological perspective, 
this might allow one to discriminate between various models and 
approximations.

There are several interesting applications of the approximation studied 
here that may be explored in future work.  One may consider the 
generalization to finite quark masses in order to compare more directly 
with lattice results.  Also unquenching effects, shown to be important 
in the lattice study of Ref.~\cite{D'Elia:2011zu}, might be looked at 
(quark loop effects were already studied in the absence of the magnetic 
field for this type of interaction in Ref.~\cite{Fischer:2005en}).  
Further, it would be interesting to study the case of finite temperature 
and chemical potential to gain insight into the QCD phase diagram.

\begin{acknowledgments}
The authors would like to thank M.~D'Elia for useful correspondence, 
in particular concerning the comparison to lattice data.  This work has 
been supported in part by the Deutsche Forschungsgemeinschaft (DFG) under 
contract no. DFG-Re856-9/1.
\end{acknowledgments}
\appendix
\section{\label{app:herm}Hermite functions and polynomials}
Let us introduce the Hermite functions, $\psi_n(x)$.  They obey the 
following differential equation ($n=0,1,2,\ldots$)
\be
\psi_n''(x)+(2n+1-x^2)\psi_n(x)=0
\ee
and form an orthonormal and complete set:
\be
\int_{-\infty}^{\infty}dx\,\psi_n(x)\psi_m(x)=\de_{nm},\;\;\;\;
\sum_{n=0}^\infty\psi_n(x)\psi_n(y)=\de(x-y).
\ee
The following recursion relations prove invaluable:
\be
\psi_n'(x)+x\psi_n(x)=\sqrt{2n}\psi_{n-1}(x),\;\;\;\;
\psi_n'(x)-x\psi_n(x)=-\sqrt{2(n+1)}\psi_{n+1}(x).
\ee
It is useful to define $\psi_{-1}(x)=0$ explicitly.  The Hermite 
functions are related to the more usual Hermite polynomials, $H_n$, 
(and parabolic cylinder functions) by
\be
\psi_n(x)=(2^nn!\sqrt{\pi})^{-1/2}e^{-x^2/2}H_n(x),\;\;\;\;
H_0(x)=1,\;\;H_1(x)=2x,\;\;H_{n+1}(x)=2xH_n(x)-H_n'(x).
\ee
The generating function for the Hermite functions is given via the identity
\be
\exp{\left\{-t^2+2tx\right\}}=\sum_{k=0}^{\infty}\frac{t^k}{k!}H_k(x).
\ee
Because of the Gaussian nature of the generating function, the Fourier 
transform of the Hermite polynomials may be performed and
\be
e^{-x^2/2}H_n(x)=(-\imath)^n\sqrt{2\pi}\int_{-\infty}^{\infty}
\frac{dp}{2\pi}e^{\imath px}e^{-p^2/2}H_n(p).
\ee
Notice the following identity \cite{grad}:
\be
\int_{-\infty}^{\infty}dx\,e^{-x^2}H_m(x+y)H_n(x+z)
=2^n\sqrt{\pi}m!z^{n-m}L_m^{n-m}(-2yz)\;\;(m\leq n),
\ee
where $L_n^\al$ is a Laguerre polynomial ($L_n^0=L_n$ and $L_{-1}=0$).  
Further,
\be
\sum_{n=0}^{\infty}z^nL_{n}^{\al}(x)=\frac{1}{(1-z)^{1+\al}}
\exp{\left\{\frac{xz}{z-1}\right\}}.
\ee

\section{\label{app:gapeqn}Explicit form of the gap equation}
For completeness, we present the explicit form for the gap equation used 
in numerical work.  The expressions are derived as follows.  The 
nonperturbative decompositions for the proper two-point function $\G$, 
\eq{eq:invprop}, and the propagator $S$, \eq{eq:prop}, both have the 
form presented in \eq{eq:anz0} and are inserted into the gap equation, 
\eq{eq:gap}, along with either of the interaction forms, \eq{eq:int}.  
The Schwinger phase is simply an overall factor.  The various Dirac 
components are then projected to generate a set of scalar equations.  
Finally, a Wick rotation to Euclidean space is performed.  In the 
presence of the magnetic field, the dressing functions 
$\hat{A}$-$\hat{E}$ are functions of two variables, $p_l^2=p_3^2+p_4^2$ 
and $p_t^2=p_1^2+p_2^2$ (in Euclidean space, $p^2=p_l^2+p_t^2$).  
The explicit equations for a chiral quark ($m=0$) then read ($q=k-p$)
\bea
\hat{A}(p_l^2,p_t^2)&=&1-\frac{dC_F}{(2\pi)^2}\int d^4k
\frac{X\exp{\left\{-\frac{q^2}{\w^2}\right\}}}{\De_E}
\left\{N_AK_{AA}+N_CK_{AC}-N_EK_{AE}\right\},\nonumber\\
\hat{B}(p_l^2,p_t^2)&=&-\frac{dC_F}{(2\pi)^2}\int d^4k
\frac{X\exp{\left\{-\frac{q^2}{\w^2}\right\}}}{\De_E}
\left\{N_BK_{BB}+N_DK_{BD}\right\},\nonumber\\
\hat{C}(p_l^2,p_t^2)&=&1-\frac{dC_F}{(2\pi)^2}\int d^4k
\frac{X\exp{\left\{-\frac{q^2}{\w^2}\right\}}}{\De_E}
\left\{N_AK_{AC}+N_CK_{AA}-N_EK_{AE}\right\},\nonumber\\
\hat{D}(p_l^2,p_t^2)&=&-\frac{dC_F}{(2\pi)^2}\int d^4k
\frac{X\exp{\left\{-\frac{q^2}{\w^2}\right\}}}{\De_E}
\left\{N_BK_{BD}+N_DK_{BB}\right\},\nonumber\\
\hat{E}(p_l^2,p_t^2)&=&1-\frac{dC_F}{(2\pi)^2}\int d^4k
\frac{X\exp{\left\{-\frac{q^2}{\w^2}\right\}}}{\De_E}
\left\{(N_A+N_C)K_{EA}-N_EK_{EE}\right\}.
\eea
In the above,
\be
X=\left\{\begin{array}{cc}\frac{q^2}{\w^2},
&\;\mbox{type I}\\1,&\;\mbox{type II}\end{array}\right.
\ee
is the factor that distinguishes between the two types of interaction.  
The propagator denominator factor occurring in the integrals above is 
given by
\be
\De_E(k_l^2,k_t^2)=k_l^2\hat{A}\hat{C}+k_t^2\hat{E}^2+\hat{B}\hat{D}
\ee
where all dressing functions are evaluated with the argument 
$(k_l^2,k_t^2)$.  The various combinations of propagator numerator 
factors (with the same arguments as for $\De_E$) are
\bea
N_A(k_l^2,k_t^2)&=&\hat{C}-\frac{h\hat{E}^2\hat{C}}{\De_E}
+\frac{\hat{D}(\hat{A}\hat{D}-\hat{B}\hat{C})}{\De_E},\nonumber\\
N_B(k_l^2,k_t^2)&=&\hat{D}-\frac{h\hat{E}^2\hat{D}}{\De_E}
-\frac{k_l^2\hat{C}(\hat{A}\hat{D}-\hat{B}\hat{C})}{\De_E},\nonumber\\
N_C(k_l^2,k_t^2)&=&\hat{A}+\frac{h\hat{E}^2\hat{A}}{\De_E}
-\frac{\hat{B}(\hat{A}\hat{D}-\hat{B}\hat{C})}{\De_E},\nonumber\\
N_D(k_l^2,k_t^2)&=&\hat{B}+\frac{h\hat{E}^2\hat{B}}{\De_E}
+\frac{k_l^2\hat{A}(\hat{A}\hat{D}-\hat{B}\hat{C})}{\De_E},\nonumber\\
N_E(k_l^2,k_t^2)&=&\hat{E}.
\eea
Finally, the integration kernels read
\bea
K_{AA}&=&\frac{1}{q^2p_l^2}
\left[q_l^2(p_l\cdot k_l)-2(p_l\cdot q_l)(k_l\cdot q_l)\right],
\nonumber\\
K_{AC}&=&\frac{1}{q^2p_l^2}
\left[-q^2(p_l\cdot k_l)-q_l^2(p_l\cdot k_l)\right],\nonumber\\
K_{AE}&=&\frac{1}{q^2p_l^2}
\left[2(p_l\cdot q_l)(k_t\cdot q_t)\right],\nonumber\\
K_{BB}&=&\frac{q_l^2}{q^2}-2,\nonumber\\
K_{BD}&=&-1-\frac{q_l^2}{q^2},\nonumber\\
K_{EA}&=&-\frac{(p_t\cdot q_t)(k_l\cdot q_l)}{q^2p_t^2},\nonumber\\
K_{EE}&=&\frac{(p_t\cdot k_t)}{p_t^2}
+2\frac{(p_t\cdot q_t)(k_t\cdot q_t)}{q^2p_t^2}.
\eea


\end{document}